\documentclass[a4paper,fleqn]{article}
\usepackage[UKenglish]{babel} %francais, polish, spanish, ...
\usepackage[T1]{fontenc}
\usepackage[ansinew]{inputenc}
\usepackage{lmodern} %Type1-font for non-english texts and characters
\usepackage{graphicx} %%For loading graphic files
\usepackage{tabularx}
\usepackage{subcaption}
\usepackage{diagbox}
\usepackage{enumitem}
\usepackage{booktabs}
\usepackage{appendix}
\usepackage{setspace}
\usepackage{natbib} % bibliography

\usepackage{amsmath}
\usepackage{amsthm}
\usepackage{amsfonts}
\usepackage{amssymb}
\usepackage{mathrsfs}
\usepackage{pbox}
\usepackage{textcomp}

\title{Linear transient growth of particulate pipe flows}
\author{Anthony Rouquier, Alban Poth\'erat and Chris C.T. Pringle}

\begin{document}

\maketitle

\begin{abstract}
We tackle the question of whether the presence of particles in a pipe flow can influence the linear 
transient growth of infinitesimal perturbations, in view of better understanding the behaviour of 
particulate pipe flows in regimes of transition to turbulence. The problem is tackled numerically by 
means of a simple model where particles are modelled as a second fluid, that interacts with the fluid 
phase through a two-way Stokes drag only.
The transient growth is found to be enhanced by the presence of particles, especially so if the 
particles are localised at a specific radial location of the pipe. At the same time, the mechanisms of 
transient growth themselves remain those of non-particulate flows.  
The effect is maximised for particles of intermediate size (somewhere between the ballistic limit of 
very light particles and the point where they are too heavy to be influenced by the flow). Most 
remarkably, the Segr\'e-Silberberg radius (around 2/3 of the pipe radius), where particles naturally 
cluster in laminar flows, turns out to be close to the optimal location to enhance the transient growth.
\end{abstract}

\section{Introduction}

This paper focuses on the topic of transition to turbulence for particulate flows. 
Transition to turbulence has been a widely studied topic since 
Reynolds first documented the phenomenon experimentally \cite{reynolds1883experimental,kerswell2005recent}. 
While much of the research on the topic has focused on single phase flows, 
there is a growing interest for particulate flows, due to their many applications. 
Examples range from the precise determination of the volume fraction of oil in the 
oil-water-sand-gas mixture that is extracted from offshore wells, to needs in the food processing 
industry \citep{ismail2005_fmi}, and flows of molten metal carrying impurities during recycling processes 
\citep{kolesnikov2011_mmtb}. 
% They require dedicated flow metering technologies, most 
% of which rely on \emph{a priori} knowledge of the nature of the pipe flow, 
% in particular whether it is turbulent or not \cite{wang2014_fmi}. 

Transition to turbulence is, even for single phase flows, a complex problem. In the case of the pipe flow, 
there is no clearly defined critical Reynolds number. 
The problem is even more complex in the case of the particulate flow, due to the large number of parameters to consider 
for the solid phase. 
To the complexity of the dynamics involved is added the inherent difficulty of considering,
theoretically or numerically, a large number of independent objects. Nonetheless, 
experimental \cite{segre1962behaviour,matas2003transition} and theoretical 
\cite{saffman1962stability,asmolov1999inertial,klinkenberg2011modal} knowledge has been amassed  
on the topic of transition to turbulence in particulate pipe flows.
For the particles influence on the pipe flow stability in particular, the effect on the transition to turbulence depends 
non-trivially on the size and volume fraction of the particles. 
Matas observed that transition occurs at lower flow rates after the addition of particles
for small particles, while the effect is reversed for large particles.
In a similar fashion, particles tend to have a destabilising effect on the pipe flow stability 
for low particle volume concentrations but a stabilising one for high volume concentrations \cite{matas2003transition}. 
Numerical simulations based on accurate modelling of individual solid particles recovered this phenomenology for pipe flows 
\citep{yu2013numerical}. 

The present knowledge on this topic is mostly empirical and there is a need for a better understanding of the 
underlying mechanisms underlying the transition to turbulence of particulate pipe flows.  
A previous study of the linear stability of particulate pipe flows uncovered a mechanism for
instability \cite{rouquier2018instability}. 
However, even when the flow is stable to infinitesimal disturbances,  
interactions between disturbances and the underlying flow can lead to large distortions of the 
base flow due to the non-normality of the linearised equations. 
Perturbations can experience large growth at finite time
\citep{waleffe1995transition,bergstrom1993optimal}, a phenomenon generally referred to as \textit{transient growth}. 
This paper aims to further this understanding using linear transient growth analysis 
in order to study the flow behaviour below the critical Reynolds number. 
% 
% 
% We study, within the framework of this two-fluid model, the particulate flow linear transient 
% growth as a function of the Reynolds number, the particles size and for
% homogeneous and nonhomogeneous particle distribution.
% 
% 
% This work focuses on the solid phase's influence on the pipe flow stability, in particular its linear transient growth. 
% The particles behaviour is expected to be non-trivial, 
% with the solid phase having a significant effect on the fluid. 
% Consequently, the model used to described the particulate flows needs to be two-way coupled in order 
% to take into account not just the effect of the particles on the fluid phase, but also the effect of 
% the fluid on the particles. 
% Less sophisticated models, for example dusty gas models \citep{marble1970dynamics} 
% or Equilibrium Eulerian models \citep{ferry2001fast,ferry2003locally}, do not contain enough of the 
% physical model characteristics to study the flow stability.  

% 
The paper starts by an introduction of the two-fluids model used  and the assumptions it relies on,
the details of the variational method used to obtain the transient growth as well as the numerical methods used (section \ref{sec:model}). 
We then consider the envelope of the optimal gain as a function of the time in section \ref{sec:envelope}. 
Section \ref{sec:homog} focuses on the effect of the Reynolds number and particle size on 
the optimal gain for homogeneously distributed particles.
It is expanded for the case of a nonhomogeneous particle distribution 
in section \ref{sec:inhomog}. 
Finally, the topology of the optimal perturbations is studied in section \ref{sec:topology}.

\section{Model and governing equations}\label{sec:model}
The complexity of particulate flows means they are usually studied through modelling assumptions and approximations  
in order to simplify the problem while keeping as much of the underlying dynamics as possible. 
In general, a trade-off has to be struck between how accurately the model represents the particulate flow and its complexity.
Models with an accurate particle description, 
such as fully Lagrangian models \citep{hu2006multi,sakai2012lagrangian,sun2013three}
and immersed boundary methods \citep{glowinski1999distributed,prosperetti2001physalis,uhlmann2005immersed}, 
suffer from a high computational cost.
% , they are ill-suited 
% to our needs; namely, to investigate the effect of particles in a more general way, 
% for a wide range of fluid and particles characteristics. 
% Moreover, the use of a individually defined particles is not adapted to the study of linear transient growth. 
% Indeed, any results depends on the precise initial conditions, and there is an almost infinite 
% number of possible initial particle position combinations.
In order to avoid the computational cost incurred when accounting for particles as individual solids, 
here we describe the particulate flow using the `two-fluid' model first derived in \cite{saffman1962stability}. 
The fluid and solid phases are treated as two inter-penetrating media, with 
the particles being described as a continuous field rather than as discrete entities. 
It is a two-way coupled model, taking into account the feedback of the solid phase on the fluid.
On the other hand, particle-particle interactions such as collisions 
or clustering, as well as the deflection of the flow around the particles, are neglected. 
The two-fluid model is therefore valid for lower concentrations and in the limit where particles are 
sufficiently smaller than the characteristic scale of the flow. 
This model has been used in the context of channel flow \cite{klinkenberg2011modal,boronin12} and boundary layer flow \cite{boronin2014modal}.
\subsection{Two-fluid model}

We consider the flow of a fluid of density $\rho_f$ and dynamic viscosity $\mu$ through a straight pipe with a constant circular 
cross-section of radius $r_0$ and driven by a constant pressure gradient. The fluid carries spherical particles of radius $a$. 

The particles are treated as a continuous field with a spatially varying density $N$. 
Their motion is coupled to the fluid via a Stokes drag force, $\mathbf S_d=6\pi a\mu |\mathbf{u_p}-\mathbf{u}|$, 
where $\mathbf{u}$ and $\mathbf{u_p}$ are the fluid and particle velocities respectively. 
When working with an averaging method, one has to ensure that 
the system of equations is closed. 
%\cite{jackson2000dynamics} laid out a general framework for averaged equations governing particulate flows. 
%In introducing a partition of the forces exerted on particles
%between the buoyancy forces and other forces, the author shows that a correction to the 
%the pressure term is needed to properly account for the direct and indirect effects of 
%buoyancy. % to be modified to account for the pressure term. 
If only the Stokes drag is considered, no specific correction is required \cite{jackson2000dynamics}. 
The Stokes drag force is proportional to $a$. 
On the other hand, other forces commonly considered (such as virtual mass force, buoyancy, Magnus force, Saffman force and the 
Basset history force) are all quadratic or above in particle radius and so can in general be neglected. 
The Stokes force, by contrast, may become significant if the background shear is 
large rather than $O(1)$ \citep{boronin2008stability}.  
Similarly, buoyancy is proportional to $\rho_f a^3$, regardless of which exact definition is chosen, 
and becomes vanishingly smaller than the Stokes drag $S_d=6\pi a\mu |\mathbf{u_p}-\mathbf{u}|$ 
in the limit $a\rightarrow0$. 
More details on the relevance of the drag-only, two-fluid model used in this paper can be found in \cite{rouquier2018instability}.

We use, the standard cylindrical set of coordinates $(r,\theta,z)$ aligned with the pipe, 
with the respective velocity components: $\mathbf{u}=(u_r,u_{\theta},u_z)$ and $\mathbf{u_p}=(u_{pr},u_{p \theta},u_{pz})$ for the fluid and particulate phases. 
Where relevant, we distinguish quantities associated with the particles from 
those associated with the fluid by means of a subscript $p$. 
The fluid velocity is described using the standard Navier-Stokes set of equations to which a Stokes drag force is added 
to account for the interaction between fluid and solid phases. 
The solid phase is characterised by the conservation of the particles momentum and density. 
Nondimensionalising by the centreline velocity, $U_0$, the pipe radius, $r_0$, and the fluid density $\rho_f$ yields the following set of governing equations:

\begin{gather}
\label{adi1}
\frac{\partial \mathbf{u}}{\partial t} = -\nabla p \, - (\mathbf{u} \cdot \nabla ) \mathbf{u} \, + \frac{1}{Re} \nabla^2 \mathbf{u} \,
+ \frac{f N}{S Re} (\mathbf{u_p} -\mathbf{u} ) \; ,  \\[1.0em]
\label{adi2}
\frac{\partial \mathbf{u_p}}{\partial t} = N (\mathbf{u_p} \cdot \nabla) \mathbf{u_p} \, 
+ \frac{1}{S Re} ( \mathbf{u} -\mathbf{u_p} ) \; ,  \\[1.0em]
\label{adii3}
\frac{\partial N}{\partial t} = - \nabla \cdot (N \mathbf{u_p}) \; ,  \\[1.0em]
\label{adi4}
\nabla \cdot \mathbf{u} = 0 \; ,
\end{gather}
where $p$ is the flow pressure and $N$ the local particle concentration.
This system is governed by three non-dimensional parameters: the Reynolds numbers $Re=U_0r_0\rho_f/\mu$, 
the aforementioned Stokes number, which expresses a dimensionless relaxation time 
$S=2a^2\rho_p/9r_0^2\rho_f$ and the mass concentration $f=m_p/m_f$, corresponding to the ratio between the 
particles and fluid mass over the entire pipe.
$N$ is normalised such that $\int N \, dV=1$. 
For a given position $\mathbf{x}$, $N(\mathbf{x}) >1$ implies that the local concentration of particles 
is higher than the pipe average.

These equations satisfy an impermeable and no-slip boundary condition for the fluid 
\begin{equation}
\mathbf{u}\vert_{r=1}=0 \label{eqn:bc_f}  \; ,
\end{equation}
and a no penetration boundary condition for the radial particle velocity
\begin{equation}
u_p\vert_{r=1}=0. \label{eqn:bc_p} \; .
\end{equation}

The stability of the flow is studied through the addition of a small perturbation to the steady solution, 
$\mathbf{U}=\mathbf{U}_p=(1-r^2)\hat{\mathbf{z}}$ :
\[ \mathbf{u}=\mathbf{U}+\mathbf{u}' , \; \; \mathbf{u_p} = \mathbf{U} + \mathbf{u_p}', \; \; p = P + p' , \; \;  N = N_0+N' \, , \]
where $N_0$ is the average local particle concentration. 
Linearising equations (\ref{adi1}) - (\ref{adi4}) around this base state yields:
\begin{gather}
\label{lin1}
\partial_t \mathbf{u}' = -\nabla p' \; -\mathbf{U} \cdot \nabla \mathbf{u}' 
\; -\mathbf{u}' \cdot \nabla \mathbf{U} \, + \frac{1}{Re} \nabla^2 \mathbf{u'} \, + \frac{f N_0}{S Re } \, (\mathbf{u_p'} -\mathbf{u'}) 
\; ,  \\[1.0em]
\label{lin2}
\partial_t \mathbf{u_p}' = -\mathbf{u_p}'\cdot \nabla \mathbf{U} \, - \mathbf{U} \cdot \nabla \mathbf{u_p}' \, + 
\frac{1}{S Re} \,( \mathbf{u'} - \mathbf{u_p'} ) \; ,  \\[1.0em]
\label{lin3}
\partial_t N = - N_0 \nabla \cdot \mathbf{u_p'} - \mathbf{u_p'} \cdot \nabla N_0 - \mathbf{U} \cdot \nabla N' \; ,  \\[1.0em]
\label{lin4}
\nabla \cdot \mathbf{u'} = 0 \; \text{.}
\end{gather}
The boundary conditions for the perturbations $\mathbf{u'}$ and $\mathbf{u_p'}$ are the same as for the full flow, $\mathbf{u}$ and $\mathbf{u_p}$ .
From here on the primes are dropped for the sake of readability.

The gain corresponds to the ratio between the maximal energy a perturbation 
% which evolves according to the linearised equations of motion 
can have at a time $T$ and the perturbation initial energy:
\begin{equation}
G(T,Re) = \max\limits_{\mathbf{u}(0)} \frac{E(\mathbf{u}(T))}{E(\mathbf{u}(0))} \; .
\end{equation}
The perturbation $\mathbf{u}(0)$ is the one causing the largest amount of 
growth, and is often referred to as the optimal disturbance. 
By optimising $G(T,Re)$ over $T$, one can find the maximum possible 
gain, or \textit{optimal gain}, at a given Reynolds number.  
This paper focuses on the optimal gain and the associated time of occurrence. 
A variational method approach, adapted from the single phase flow problem \cite{pringle2010using}, is used to solve this optimisation problem. 
The problem described by equations (\ref{lin1})-(\ref{lin4}) can be characterised with the following functional $\mathcal{L}$: 

\begin{align}
\label{func1}
\mathcal{L} = & \left\langle \frac{1}{2} \Big(m_f \mathbf{u}^2(T) + m_p \mathbf{u^2_p}(T) \Big) \right\rangle - 
\lambda \left[ \left\langle \frac{1}{2} \Big(m_f \mathbf{u^2}(0) + m_p \mathbf{u^2_p}(0) \Big) - E_0 \right\rangle \right] \nonumber \\[0.9em]
 & - \, \int_0^T \! \left\langle \boldsymbol{\Upsilon}\, \cdot \, 
 \Big( \partial_t \mathbf{u} + \nabla p \; + \mathbf{U} \cdot \nabla \mathbf{u} \; -\mathbf{u} \cdot \nabla \mathbf{U} \, - 
\frac{1}{Re} \nabla^2 \mathbf{u} \, - \frac{f N_0}{S Re} \, (\mathbf{u_p} -\mathbf{u}) \Big) \right\rangle \mathrm{d}t \, \nonumber \\[0.9em] 
 & -\int_0^T \! \left\langle \boldsymbol{\Upsilon_p} \, \cdot \, 
  \Big( \partial_t \mathbf{u_p} + \mathbf{u_p} \cdot \nabla \mathbf{U} \, + \mathbf{U} \cdot \nabla \mathbf{u_p} \, 
  - \frac{1}{S Re} \,( \mathbf{u} - \mathbf{u_p} ) \Big) 
  \right\rangle  \mathrm{d}t  \nonumber \\[0.9em]
 & - \, \int_0^T \! \left\langle \Pi \, \cdot \, \nabla \cdot \mathbf{u} \right\rangle \mathrm{d}t \,
 - \int_0^T \!  \left\langle  \Gamma \, \cdot \,  (
 \partial_t N + N_0 \nabla \cdot \mathbf{u_p} + \mathbf{u_p} \cdot \nabla N_0 + \mathbf{U} \cdot \nabla N )
 \right\rangle \, \mathrm{d}t \; ,
\end{align}
where $ \lambda $, $\boldsymbol{\Upsilon} $, $\boldsymbol{\Upsilon_p} $, $\Gamma$ 
and $\Pi$ are the Lagrange multipliers enforcing the constraints of the problem: 
$ \lambda $ enforces that the energy is fixed, $\boldsymbol{\Upsilon} $, $\boldsymbol{\Upsilon_p} $ and $\Gamma$ 
enforce that Equations (\ref{lin1}) and (\ref{lin2}) hold true over $ t \in [0, T] $, 
$\Pi$ and $\Gamma$ enforces the incompressibility of the flow and
the conservation of the total number of particles.
The brackets represent a normalised volume integral over the pipe, given any function $f$: 
$ \; \left\langle f \right\rangle = \int \!  f \, \mathrm{d}V / V_p $ with $V_p$ the pipe volume. 

Finding the initial perturbation that will maximise energy growth is equivalent to maximising $\mathcal{L}$, done
here through finding the root of its variational derivative $\delta \mathcal{L}$. 
By reordering $\delta \mathcal{L}$, one can obtain the adjoint system of equations
of our problem, with an additional set of conditions. 
The adjoint system of equation is: 
 \begin{gather}
 \label{Adj1}
\partial_t \boldsymbol{\Upsilon} = - \, \mathbf{U} \cdot \nabla \boldsymbol{\Upsilon} \; + \, \boldsymbol{\Upsilon} \cdot \nabla \mathbf{U} \, 
- \nabla \boldsymbol{\Pi }
-\frac{1}{Re} \nabla^2 \boldsymbol{\Upsilon}\,+ \frac{f N_0}{S Re} \, \boldsymbol{\Upsilon}\,-\,\frac{1}{S Re} \,\boldsymbol{\Upsilon_p }  \; , \\[1.0em]
 \label{Adj2}
\partial_t \boldsymbol{\Upsilon_p} = - \, \mathbf{U} \cdot \nabla \boldsymbol{\Upsilon_p} \, + \, \boldsymbol{\Upsilon_p} \cdot \nabla \mathbf{U} 
- N_0 \, \nabla \Gamma - \frac{f N_0}{S Re} \, \boldsymbol{\Upsilon} \, + \, \frac{1}{S Re} \boldsymbol{\Upsilon_p} \; ,  \\[1.0em]
 \label{Adj3}
\partial_t \Gamma = - \, \mathbf{U} \cdot \nabla \Gamma - \mathbf{u_p} \cdot \nabla \Gamma \; ,  \\[1.0em]
 \label{Adj4}
 \nabla \cdot \boldsymbol{\Upsilon} = 0 \; .
\end{gather}
where $\boldsymbol{\Upsilon}$ and $\boldsymbol{\Upsilon_p}$ are 
the adjoint fluid and particles velocities respectively; $\Gamma$ is 
 the adjoint particle local concentration while $\Pi$ is the adjoint pressure. 
The adjoint equations must be true for $\delta \mathcal{L}$ to be equal to $0$.
Enforcing $\delta \mathcal{L} = 0$ yields another set of conditions: 
\begin{gather}
 \label{Adj5}
 \mathbf{u}(T) = \boldsymbol{\Upsilon}(T) \quad , \quad \mathbf{u_p}(T) = \boldsymbol{\Upsilon_p}(T) \; ,  \\[1.0em]
 \label{Adj6}
\lambda \mathbf{u}(0) - \boldsymbol{\Upsilon}(0) = 0 \quad , \quad \lambda \mathbf{u_p}(0) - \boldsymbol{\Upsilon_p}(0) = 0 \; .
\end{gather}
 
%A full derivation of the adjoint equations and additional conditions is given in the Appendix. 
In this paper we consider homogeneous and nonhomogeneous particle distributions. 
In the case of a homogeneous particle distribution, $N_0$ is held constant spatially.
However, particles are not necessarily uniformly distributed in practice. 
In particular, they 
tend to aggregate in the radial direction, around $r=0.6-0.8$ \citep{segre1962behaviour, matas2004inertial}. 
We parametrise this phenomenon by mean of a particle distribution of the form  
\begin{equation}
N_0(r)=\tilde{N}\exp\{-(r-r_d)^2/2\sigma^2\},
\end{equation}
with $\tilde{N}$ chosen such that $\int_0^1 N_0(r)rdr=1$. 
The distribution is then, in the radial direction, a Gaussian centred around radius $r_d$ with a standard deviation $\sigma$.  
$N_0$ is still homogeneous in the axial and azimuthal directions.
 
A point of note is that, as opposed to the single phase pipe flow which 
is well-known to be linearly stable, particulate flow can, within our theoretical framework, be linearly unstable 
in the case of nonhomogeneous particle distributions \cite{rouquier2018instability}. 
However, only linearly stable cases are considered in this work.

\subsection{Iterative variational method}

We use an iterative procedure to minimise $\delta \mathcal{L}$, 
akin to the one used in \cite{pringle2012minimal}.
Initially, a first guess of the initial perturbation is made for the fluid velocity  $\mathbf{u}^{(0)}(t=0) = \mathbf{u_0}^{(0)}$ and 
 the particles velocity $\mathbf{u_p}^{(0)}(t=0) = \mathbf{u_{p0}}^{(0)}$. 
The initial perturbations of the fluid and solid phases for iteration $(i+1)$ are: 
 \begin{equation}
 \mathbf{u}^{(i+1)}(0) = \mathbf{u}^{(i)}(0) + \epsilon ( \lambda \mathbf{u}^{(i)}(0) - \boldsymbol{\Upsilon}^{(i)}(0) ) %\; ,
 \end{equation}
for the fluid velocity, and:
 \begin{equation}
\mathbf{u_p}^{(i+1)}(0) = \mathbf{u_p}^{(i)}(0) + \epsilon_p (\lambda_p \mathbf{u_p}^{(i)}(0) - \boldsymbol{\Upsilon_p}^{(i)}(0)) \; 
 \end{equation}
for the particle velocity. It entails that $\boldsymbol{\Upsilon}(0)$ and $\boldsymbol{\Upsilon_p}(0)$ have to be computed for each iteration. 
To that effect, the iteration process is as follows:
\begin{itemize}
\item At the $i$-th iteration, equations (\ref{lin1})-(\ref{lin4}) are advanced from $t=0$ until a target time $t=T$ is reached in order 
to obtain $\mathbf{u}^{(i)}(T)$ and $\mathbf{u_p}^{(i)}(T)$.  
\item $\boldsymbol{\Upsilon}^{(i)}(T)$ and $\boldsymbol{\Upsilon_p}^{(i)}(T) $ are then computed using conditions (\ref{Adj5}).
\item The adjoint system of Equations (\ref{Adj1})- (\ref{Adj4}) is run from $t=T$ to $t=0$ to find $\boldsymbol{\Upsilon}^{(i)}(0)$ and $\boldsymbol{\Upsilon_p}^{(ii)}(0)$ 
\item The final conditions 
 \begin{equation}
 \frac{\partial \mathcal{L}}{\partial\mathbf{u_0}} = - \lambda \mathbf{u_0} - \boldsymbol{\Upsilon} \; , \quad 
 \frac{\partial \mathcal{L}}{\partial\mathbf{u_{p0}}} = - \lambda_p \mathbf{u_{p0}} - \boldsymbol{\Upsilon_p} \; 
 \end{equation}
 give the gradient in $\mathbf{u_0}$ and $\mathbf{u_{p0}}$ and the initial conditions are updated as
 \begin{equation}
 \mathbf{u}^{(i+1)}(0) = \mathbf{u}^{(i)}(0) + \epsilon \frac{\partial \mathcal{L}}{\partial\mathbf{u_0}} \; , \quad 
 \mathbf{u_p}^{(i+1)}(0) = \mathbf{u_p}^{(i)}(0) + \epsilon \frac{\partial \mathcal{L}}{\partial\mathbf{u_{p0}}}, 
 \end{equation}
 where $\epsilon$ is the step size.
\end{itemize}
The process is repeated until the norms of $\partial \mathcal{L}/\partial\mathbf{u_0}$ and 
$\partial \mathcal{L}/\partial\mathbf{u_{p0}}$ are less than a threshold chosen for convergence.

\subsection{Computational method}

The code is derived from a standard DNS code \cite{openpipeflow}.
Temporal discretisation is done through a predictor-corrector scheme.
The spatial discretisation is done using a fourth order finite difference method in the 
radial direction and Fourier spatial discretisation with $128$ mesh points used in the azimuthal and streamwise directions.
Any field $\mathbf{g}$ can then be written as:
\begin{equation}
\mathbf{g}(r,\theta,z ,t ) = \sum_{\alpha} \sum_{m} \hat{g}(r) e^{i (\alpha z + m \theta - \omega t) } \; ,
\end{equation}
where $\alpha$ and $m$ are the wavenumbers in the streamwise and azimuthal directions respectively.

%The time unit used in this paper corresponds to the numerical time of the code \citep{openpipeflow}; 
%it is equal to $t = r_0 / U_{cl} $ where $r_0$ is the pipe radius and $U_{cl}$ is 
%the centreline velocity of the steady Hagen-Poiseuille flow.
% 
%
The numerical code has been modified in order to 
incorporate the solid phase, using a fully Eulerian method. 
First, we add a set of equations for the particles velocity, $\mathbf{u_p}$, for both the standard and adjoint problems 
(Equations (\ref{lin2}) and (\ref{Adj2}) respectively). 
Initial and boundary conditions for the particle velocity (equation \ref{eqn:bc_p}) are added as well.  
The initial fluid velocity is obtained from a previously saved state. 
%
%In the single phase linear DNS, the convergence of the problem is obtained by ensuring that $\frac{\partial \mathcal{L}}{\partial\mathbf{u_{0}}}$ tends 
%towards $0$. 
%An additional condition is added to the problem in the case of the particulate flow: 
%$ \frac{\partial \mathcal{L}}{\partial\mathbf{u_{p0}}} =0$. 
%The iterative process of the code had to be modified to compute and ensure the decay 
%of both $\frac{\partial \mathcal{L}}{\partial\mathbf{u_{0}}}$ and  $\frac{\partial \mathcal{L}}{\partial\mathbf{u_{p0}}}$.
%

%
\begin{figure}
\centering
\includegraphics[width=0.85\textwidth]{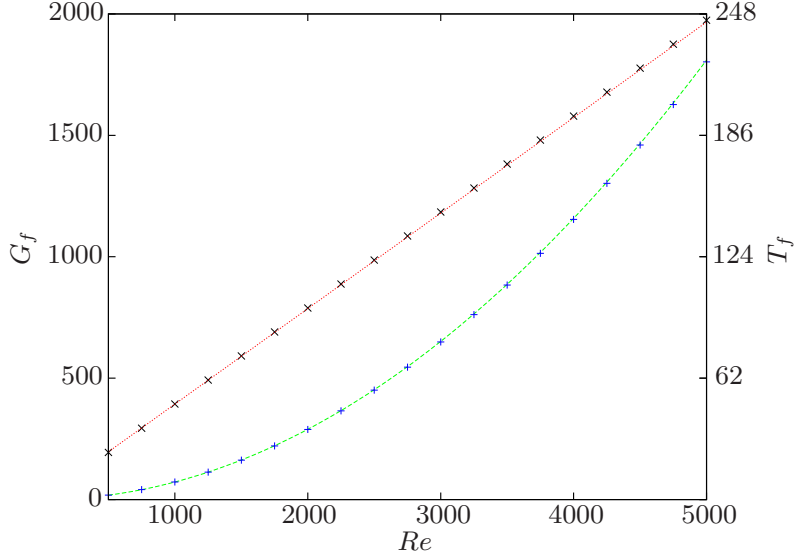}
\caption{Optimal gain $G_f$ (green) and optimal time of gain $T_f$ (red) for the single phase flow as a function of the Reynolds number. 
The points correspond to values obtained using our code. The lines corresponds to the scaling 
given in \cite{schmid2012stability}: $G_f=72.40  Re^2 \times 10^{-6}$ , $T_f=48.77 Re \times 10^{-3}$.  }
\label{fig:fl}
\end{figure}

\begin{figure}
\centering
\includegraphics[width=1\textwidth]{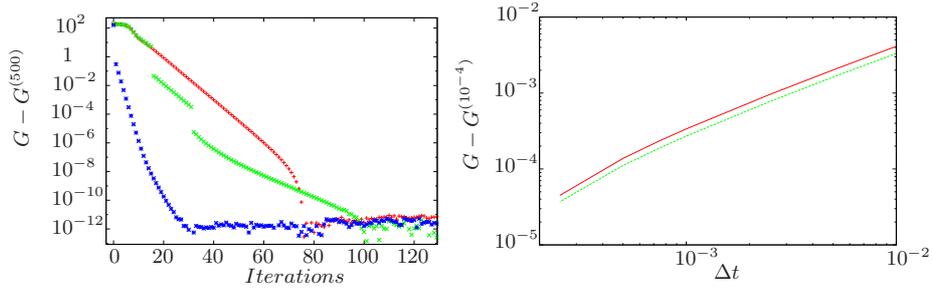}
\caption{ $S=10^{-3}$, $f=0.1$, $Re=1000$.
\textbf{Left:} Optimal gain as a function of the number of iterations $n$, within the  
optimisation process. Single phase flow (blue dots), two particulate cases are shown, where $G$ is either computed with a fixed value of $T=50$ (red dots) 
or $G$ is optimised over $T$ (green dots).
\textbf{Right:} Optimal gain as a function of the time step for single phase (red) and particulate flows (green), $S=10^{-3}$, $f=0.1$, $Re=1000$. \\
}\label{fig:conv}
\end{figure}

%\vspace{-1\baselineskip}

\subsection{Code validation and convergence}

The code has been first verified against the literature on the single phase pipe flow, 
which is simulated by fixing the particle mass concentration $f$ to $0$. 
\cite{bergstrom1992initial}  
found that the time of the peak in energy increases linearly with the Reynolds number while the 
optimal gain scales with $Re^2$ for all modes, with 
$G_f=72.40  Re^2 \times 10^{-6}$ and $T_f=48.77 Re \times 10^{-3}$ \cite{schmid2012stability}.
These scalings have been recovered with our code, as illustrated in figure \ref{fig:fl}. 
Second, the growth rate of the leading eigenvalue obtained through a linear stability analysis of the system of equations (\ref{lin1})-(\ref{lin4})
is proportional to the energy decay rate of the linear DNS at large times and therefore offers a convenient way to test the long term 
evolution of individuals modes in the DNS code. 
Table \ref{comparison1} and Table \ref{comparison2} show the leading eigenvalue found with linear stability analysis and LDNS simulation, 
for a single phase and particulate flow respectively. 
The normalised error is always below $10^{-3}$.
% , the difference also tends to be lower for lower values of $Re$ and $S$.
The difference between the linear stability analysis and linear DNS results is not increased by the addition of particles.\\

Figure \ref{fig:conv} shows the difference between the values for optimal gain obtained 
for a given number of iterations, and a fully converged value, $G^{(500)}$.
The growth $G$ is shown to converge as the process is iterated, 
reaching fully converged values after a sufficient number of iterations in the three cases considered in figure \ref{fig:conv}.
The number of iterations needed to fully converge depends on the case, 
convergence is significantly faster in the case of a single phase flow, where $30$ iterations are typically needed to reach machine precision; 
whereas in the case of particulate flows, this number varies between $80$ and $100$. 
The number of iterations to reach machine precision can also be 
decreased by choosing initial velocity profiles closer 
to the ones leading to optimal growth. \\

%In this case, the initial perturbation was random, convergence can also be accelerated by choosing a perturbation closer 
%to the optimal perturbation expected.
The optimal gain also converges as the time step decreases 
following a power law as illustrated in Figure \ref{fig:conv}.
The time step chosen in this study is, unless otherwise specified, $\Delta t =10^{-3}$, to 
obtain a good compromise between accuracy and computational cost. 
As we observe asymptotic behaviours 
for extreme values of $S$, these are less relevant, we therefore use in this work values of $S$ ranging form $10^{-4}$ to $10^{-1}$ 
as it is the region where interesting behaviour is observed. 
We chose to keep $f$ constant at $f=0.1$, as $f$ was not shown to significantly 
impacts the results found, similarly to what has been observed in the case of 
the linear stability analysis with the same model \cite{rouquier2018instability}. 
Reynolds numbers are considered up to $Re=10^4$ as the behaviour showed little change 
with variations of $Re$ and large values are less relevant within the linear approximation we 
consider.

\newpage 

\begin{center}
\begin{tabular}{c c c c c c} % | pour les encadrements verticaux et \line pour les encadrements horizontaux 
 \toprule
 $Re$ & $\alpha$ & $m$ & Eigenvalue solver & LDNS & $\epsilon$\\
 \midrule
 $1000$ & $0$ & $1$ & $-1.4682 \times 10^{-2}$ & $-1.4681 \times 10^{-2}$ & $5.5853 \times 10^{-5}$ \\
 $3000$ & $0$ & $1$ & $-4.8940 \times 10^{-3}$ & $-4.8866 \times 10^{-3}$ & $1.5121 \times 10^{-3}$ \\
 $5000$ & $0$ & $1$ & $-2.9364 \times 10^{-3}$ & $-2.9344 \times 10^{-3}$ & $6.9658 \times 10^{-4}$ \\
 $1000$ & $1$ & $0$ & $-7.0864 \times 10^{-2}$ & $-7.0898 \times 10^{-2}$ & $4.7956 \times 10^{-4}$ \\
 $3000$ & $1$ & $0$ & $-4.1276 \times 10^{-2}$ & $-4.1317 \times 10^{-2}$ & $1.0131 \times 10^{-3}$ \\
 $5000$ & $1$ & $0$ & $-3.2043 \times 10^{-2}$ & $-3.2087 \times 10^{-2}$ & $1.3604 \times 10^{-3}$ \\
 $1000$ & $1$ & $1$ & $-9.0443 \times 10^{-2}$ & $-9.0483 \times 10^{-2}$ & $4.3953 \times 10^{-4}$ \\
 $3000$ & $1$ & $1$ & $-5.1973 \times 10^{-2}$ & $-5.2018 \times 10^{-2}$ & $8.7257 \times 10^{-4}$ \\
 $5000$ & $1$ & $1$ & $-4.0200 \times 10^{-2}$ & $-4.0246 \times 10^{-2}$ & $1.1504 \times 10^{-3}$ \\
 \bottomrule
\end{tabular}
\end{center}
\vspace{-0.5\baselineskip}
\captionof{table}{Comparison of long term decay rates of linearly stable 
eigenmodes obtained from LSA (eigenvalue solver) and through our DNS code for a single phase flow. $\epsilon = \frac{\vert \omega_{lsa}-
\omega_{LDNS} \vert}{\omega_{LDNS}} $, $\Delta t = 10^{-3} $.}
\label{comparison1}

\vspace{1\baselineskip}

\begin{center}
\begin{tabular}{c c c c c c} % | pour les encadrements verticaux et \line pour les encadrements horizontaux 
 \toprule
 $S$ & $\alpha$ & $m$ & Eigenvalue solver & DNS & $\epsilon$\\
 \midrule
 $10^{-4}$ & $0$ & $1$ & $-1.4526 \times 10^{-2}$ & $-1.4526 \times 10^{-2}$ & $5.5075 \times 10^{-6} $ \\
 $10^{-3}$ & $0$ & $1$ & $-1.4536 \times 10^{-2}$ & $-1.4523 \times 10^{-2}$ & $8.3513 \times 10^{-4} $ \\
 $10^{-2}$ & $0$ & $1$ & $-1.4513 \times 10^{-2}$ & $-1.4501 \times 10^{-2}$ & $8.7025 \times 10^{-4} $ \\
 $10^{-1}$ & $0$ & $1$ & $-8.4935 \times 10^{-3}$ & $-8.4931 \times 10^{-3}$ & $4.8274 \times 10^{-5} $ \\
 $10^{-4}$ & $1$ & $0$ & $-8.9988 \times 10^{-2}$ & $-9.0029 \times 10^{-2}$ & $4.5108 \times 10^{-4} $ \\
 $10^{-3}$ & $1$ & $0$ & $-8.9981 \times 10^{-2}$ & $-8.9977 \times 10^{-2}$ & $4.7790 \times 10^{-5} $ \\
 $10^{-2}$ & $1$ & $0$ & $-8.9791 \times 10^{-2}$ & $-8.9855 \times 10^{-2}$ & $7.5478 \times 10^{-4} $ \\
 \bottomrule
\end{tabular}
\end{center}
\vspace{-0.5\baselineskip}
\captionof{table}{Comparison of long term decay rates of linearly stable 
eigenmodes obtained from LSA (eigenvalue solver) and through our DNS code for particulate flows. $\epsilon = 
\frac{\vert \omega_{lsa}-\omega_{LDNS} \vert}{\omega_{LDNS}}$, $Re=1000$, $f=0.01$, $\Delta t = 10^{-3} $. }
\label{comparison2}
\vspace{2\baselineskip}
\section{Growth envelope}\label{sec:envelope}
\begin{figure}
\centering
\includegraphics[width=1\textwidth]{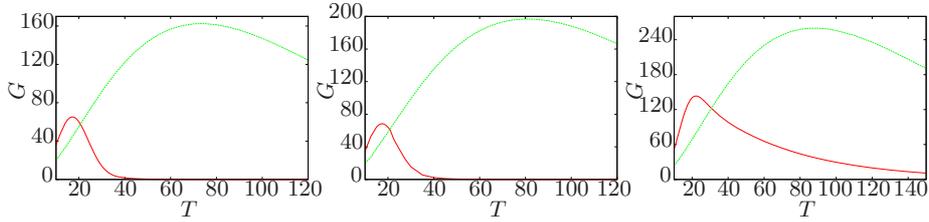}
\caption{Maximal growth as a function of the time of optimisation, $T$ with $Re=1500 $. 
From left to right : Single phase flow ;  uniform particle distribution with $S=10^{-3} $ and $f=0.1 $ ; % 
 Gaussian particle distribution with $r_d = 0.65$, $\sigma=0.104$, $S=10^{-3}$ and $f=0.1 $. 
Wavenumbers ($\alpha,m$) = ($1,1$) in red, ($\alpha,m$) = ($0,1$) in green. }
\label{env1}
\end{figure}
The value of the maximum transient growth depends on the target time chosen.
While we are mostly interested in optimising for $T$, it is still interesting to see how $G$ depends on $T$.
Figure \ref{env1} shows the growth envelope (from left to right) for a single phase flow  
and two examples of particulate flows with homogeneous and nonhomogeneous particle distribution.
The two modes showing the largest growth, ($\alpha,m$) = ($0,1$) and ($\alpha,m$) = ($1,1$), are plotted independently. 
The envelopes are of similar shape in all three cases. 
We observe two competing mechanisms for growth: 
% 
%\begin{itemize}
%\item 
at small times, below $T \approx20$ in the single phase flow case, and $T\approx25-30$ for the particulate flow for the examples shown in figure \ref{env1}, 
the mode producing the most growth is ($\alpha,m$) = ($1,1$).
The growth produced by this mode quickly decreases as the time increases.
%\item 
At larger values of $T$, the mode  producing the most growth is ($\alpha,m$) = ($0,1$).
%\end{itemize}
% 
For single phase pipe flows, ($\alpha,m$) = ($0,1$) is the mode that yields the maximal gain when optimising for the target time $T$ \cite{bergstrom1992initial}.
This is also the case for particulate flows, whether the particle distribution is homogeneous or not. 
This result, plus the similar shapes of the envelopes, 
suggests that the mechanisms producing growth are the same for single phase and particulate flows. 
The mode ($\alpha,m$) = ($1,1$) is the most affected 
by the addition of particles, especially for nonhomogeneous particle distributions as 
illustrated in figure \ref{env1}. 
However, ($\alpha,m$) = ($1,0$) is still the mode for which the gain is the strongest. 

From now on, $G$ is optimised over $T$ when studying the optimal gain and 
the value of the mass fraction $f$ is also kept constant, to $f=0.1$.

\section{Homogeneous particle distribution}\label{sec:homog}

% \begin{figure}
% \centering
% \includegraphics[height=0.35\textwidth]{growth_paper.eps}
% \caption{Ratio of growth between particulate and single phase flow as a function of $S$ for $f = 0.1$. 
% \textbf{Left:} Homogeneous particle distribution, varying Reynolds number. 
% $Re = 500$ (red), $1000$ (green), $2000$ (dotted blue), $3000$ (purple), $5000$ (blue dashed).
% \textbf{Right:} Gaussian particle distribution with $\sigma=0.1$, $Re=1000$:
% Uniform distribution (red), $r_d=0.3$ (green), $r_d=0.5$ (dotted blue), $r_d=0.6$ (purple), $r_d=0.7$ (dashed blue), $r_d=0.8$ (yellow).}
% \label{fig:growth1}
% \end{figure}

\begin{figure}
\centering
\includegraphics[width=1\textwidth]{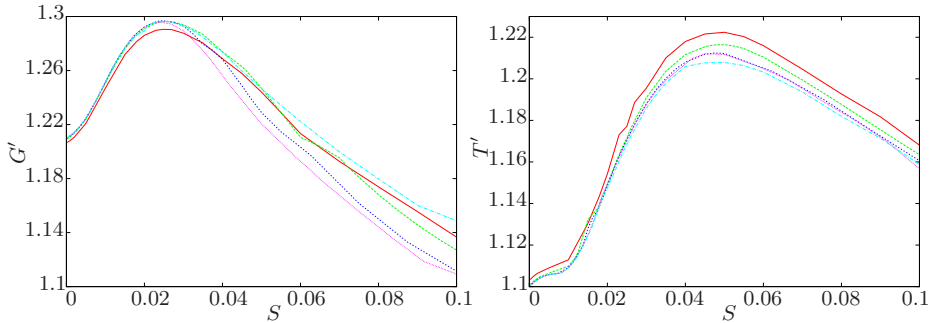}
\caption{Ratio of growth between particulate and single phase flows (at the same $Re$) as a function of $S$ for $f = 0.1$.
\textbf{Left:} Ratio of optimal gain. \textbf{Right:} Ratio of the time of maximum growth. 
$Re = 500$ (red), $1000$ (green), $2000$ (dotted blue), $3000$ (purple), $5000$ (blue dashed).}
\label{fig:uniG}
\end{figure}
We first examine the effect of adding homogeneously distributed particles in the flow.
In order to illustrate the effect of particles, we define the ratio between the growth 
for the particulate flow with a given set of parameters and the single phase flow with the same Reynolds number:
\begin{equation}
G' = \frac{G_p(Re , S,f)}{G_f(Re)}, 
\end{equation}
where $G_p$ and $G_f$, are the optimal gains for particulate and single phase flows respectively, both
 maximised over all values of $T$. A similar ratio is chosen between the times of optimal growth 
for particulate and single phase flow, 
\begin{equation}
T' = \frac{T_p(Re ,S,f)}{T_f(Re)},
\end{equation}
 where $T_p$ and $T_f$ are the target time associated to $G_p$ and $G_f$.

\subsection{Effect of the Stokes numbers on the gain}

We first consider variations of these quantities as a with the Stokes number. 
Figure \ref{fig:uniG} shows $G'$ and $T'$ as a function of $S$, 
for different values of $Re$.\\
The addition of particles increases the optimal gain for all values of $S$. 
The curves are non monotonic, with $G'$ increasing  
until it reaches a peak defined as $G'_{peak}$, for an associated Stokes number $S_G$.
For large values of $S$, the ratio $G'$ seems to decreases towards $1$. 
In the limit of $S \rightarrow \infty$, the particles are so heavy that they are effectively decoupled from the flow 
and have no effect on it.
When $S \to 0$, $G' \approx 1.21 $. 
The difference between single phase and particulate flows is due to the modification of the average density of the flow 
caused by the particles. 
With $f=0.1 $, the modified Reynolds number is $Re' = (1+f) Re= 1.1 Re $.   
Since for the single phase pipe flow, $G_f \propto Re^2 $, $G_p \propto (1+f)^2 Re^2$,  
as observed in Figure \ref{fig:uniG}. 
It follows that there is an optimal Stokes number $S$ for which the influence 
of homogeneously distributed particles on the optimal gain is greatest. 

A similar behaviour is observed for $T'$. 
For all values of $Re$ and $S$ considered, the growth is delayed for particulate flows compared to single phase flows.
As $S \to 0$, the time for which the growth is maximised increases by $10 \%$ compared to the single phase flow. 
This corresponds to the time of optimal growth for the modified Reynolds number $Re' = (1+f) Re$ since, as 
discussed in the previous section, the time for maximum growth increases linearly with $Re$.
Similarly, a peak for the time ratio $T'_{peak}$ occurs at a Stokes number $S_T$. 
The time ratio then decreases as the Stokes number continues to increase in a similar fashion as the ratio of growths. For all Reynolds numbers considered and $f=0.1$, the peak Stokes number is around 
$S_T = 5 \times 10^{-2}$. 
% $S_T$ is significantly higher than $S_G$, in addition there also is more variations as $Re$ vary. 
%
\subsection{Effect of the Reynolds number}

The Reynolds number has little incidence on $G'$, as the envelopes have a very similar shape when $Re$ is varied.  For all $Re$ considered, the curves of Figure \ref{fig:uniG} exhibit  
a peak at approximately the same Stokes number, $S_G \simeq 2.5 \times 10^{-2}$.

The value $G'_{peak}$ of the peak shows little change, varying by only $0.25\%$ for $Re$ ranging from $500$ to $5000$. 
Since $G'_{peak}$ is almost constant over the Reynolds number 
and the optimal gain for single phase flow $G_f(Re)$ scales with $Re^2$ \citep{bergstrom1992initial}, 
it follows that the optimal gain for particulate flows optimised over $S$ also 
scales with $Re^2$.
Similarly, the ratio of delays $T'_{peak}$ varies little with the Reynolds number, with variations just 
under $1\%$ for $Re$ ranging from $500$ to $5000$. 
Moreover, $T_f$ scales linearly with the Reynolds number. 
Therefore $T_p$ optimised over $S$ scales linearly with the Reynolds number as well.

\section{Inhomogeneous particle distribution}\label{sec:inhomog}

% \begin{figure}
% \centering
% \includegraphics[width=1\textwidth]{growth_paper2.eps}
% \caption{Ratio of growth between particulate and single phase flow as a function of $S$ for $f = 0.1$. 
% \textbf{Left:}  Gaussian particle distribution centred around $r_d=0.3$, $Re=500$.
% \textbf{Right:}  Gaussian particle distribution centred around $r_d=0.7$, $Re=500$.
% Uniform distribution (red), $\sigma=0.15$ (green), $\sigma=0.12$ (dotted blue), $\sigma=0.10$ (purple).}
% \label{fig:growth2}
% \end{figure}
% % 

Particles do tend to cluster in laminar pipe flows \cite{matas2003transition} 
such that considering homogeneously distributed particles is less realistic.  
Moreover, allowing for an inhomogeneous distribution dramatically 
increases the effect of the solid phase in the case of the linear stability \cite{rouquier2018instability}. 
% AP: I REMOVED THE SENTENCE BELOW BECAUSE IT IS NOT SUPPORTED BY EVIDENCE AT THIS STAGE
%On the other hand, while the growth significantly increases when using 
%inhomogeneous distribution, the mechanisms producing transient growth stays the same.

\begin{figure}
\begin{subfigure}{1\textwidth}
\centering
\includegraphics[width=1\textwidth]{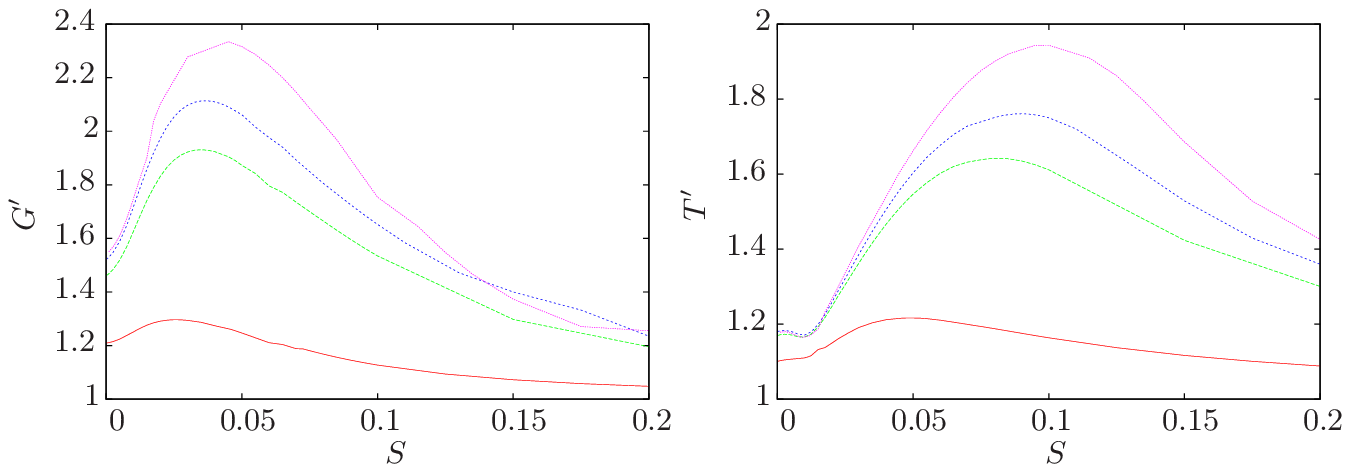}
\caption{$Re=1000$ and $r_d=0.7$. 
}
\label{fig:g1}
\end{subfigure}\hfill
\begin{subfigure}{1\textwidth}
\centering
\includegraphics[width=1\textwidth]{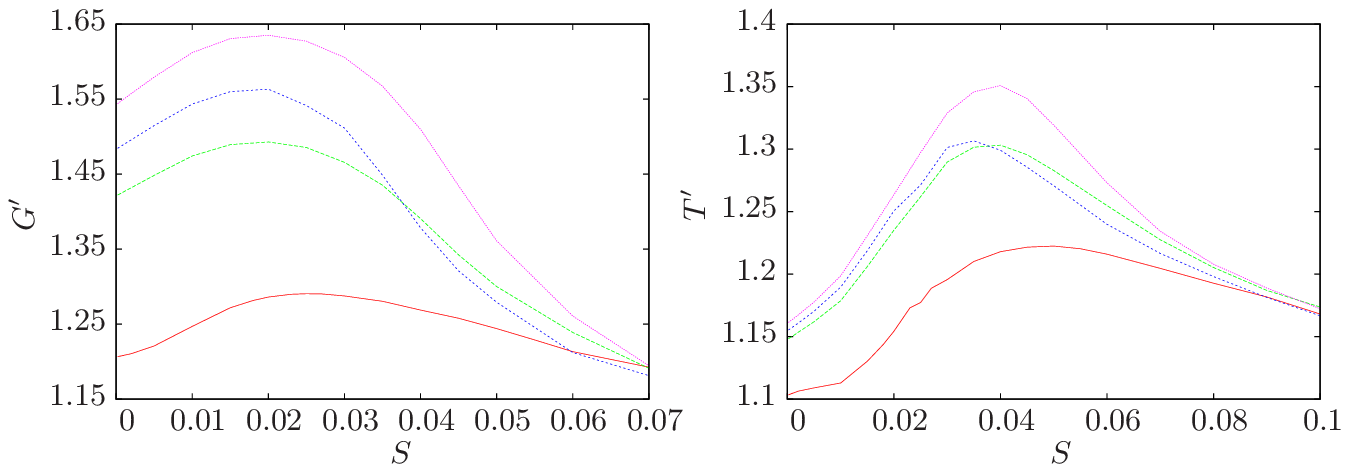}
\caption{$Re=500$ and $r_d=0.3$.
}
\label{fig:g2}
\end{subfigure}
\caption{Ratio of optimal gain $G'$ (left) 
and time of optimal gain $T'$ (right) as a function of $S$ for $f = 0.1$  in the case of a Gaussian particle distribution.
Uniform distribution (red), $\sigma=0.15$ (green), $\sigma=0.12$ (dotted blue), $\sigma=0.10$ (purple).}
\label{fig:growth2}
\end{figure}

\begin{figure}
\centering
\includegraphics[width=1\textwidth]{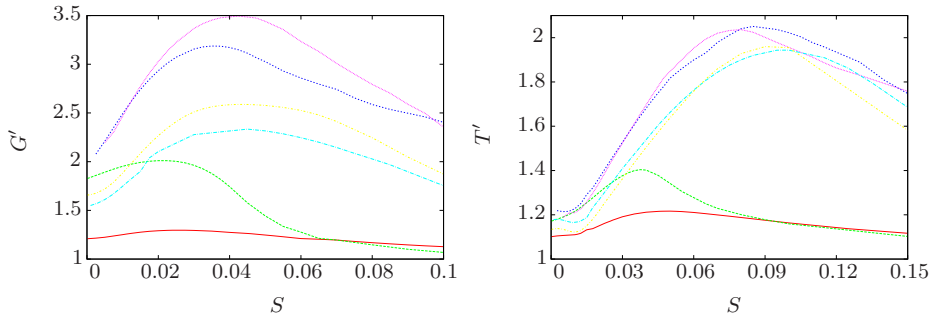}
\caption{Ratio of growth between particulate and single phase flow as a function of $S$ for $f = 0.1$ and $Re=1000$ in the case of a Gaussian particle distribution with $\sigma=0.1$. 
\textbf{Left:} Optimal gain. \textbf{Right:} Time of optimal gain. 
Uniform distribution (red), $r_d=0.3$ (green), $r_d=0.5$ (dotted blue), $r_d=0.6$ (purple), $r_d=0.7$ (dashed blue), $r_d=0.8$ (yellow).}
\label{grsd}
\end{figure}

\subsection{Influence of the distribution standard deviation}

Figure \ref{fig:growth2} shows $G'$ and $T'$ for varying values of $\sigma$, 
 centred around $r_d=0.7$ (figure \ref{fig:g1}) and $r_d=0.3$ (figure \ref{fig:g2}). 
The overall shape of the growth is the same as in the case of a homogeneous particle distribution, 
but the effect of the solid phase on the gain is significantly stronger. 
It also significantly varies with $\sigma$. 

The more the particles are concentrated, \emph{i.e.} the smaller $\sigma$, the larger both $G'$ and $T'$ are, as illustrated in figure \ref{fig:growth2}.
Varying $\sigma$ from $0.15$ to $0.10$, $G'_{peak}$ increases by $ 24\%$ and $T'_{peak}$ by $ 18\%$ for $r_d=0.7$ (table \ref{tab:inhomog}).\\
The effect of $\sigma$ is similar for $r_d=0.3$, as seen in figure \ref{fig:g2}.  
However, the values of $G'_{peak}$ and $T'_{peak}$ are noticeably smaller for equivalent values of $\sigma$.  
This indicates that the position of the particles determine the amount of transient growth as well.
% Another point of note is that, for non homogeneous particle distributions,  
% the growth ratio $G_p$ as $S \to 0$ is no longer equal to $(1+f)^2 G_f $, as it 
% was the case for homogeneously distributed particles (section \ref{sec:homog}).
However, both $G'$ and $T'$ still tends towards $1$ as $S \to \infty$ in all cases. Compared to 
homogeneous particles distribution, the values of $S$ yielding the maximal gain shifts to a larger 
value for $r_d=0.7$.  
The time at which the optimal growth occurs is delayed for Gaussian particle distributions as well. 
On the other hand these effects are reversed for $r_d=0.3$. 
% The behaviour as a function of $S$ in general is not affected by variations of either $r_d$ or $\sigma$. 
Moreover, while changing $\sigma$ affected the growth ratio, it has  
little effect on the value of $S_G$ and $S_T$ for all cases observed. 

\begin{center}
\begin{tabular}{c c c c c c} % | pour les encadrements verticaux et \line pour les encadrements horizontaux 
 \toprule
 $r_d$ & $\sigma$ & $G'_{peak}$ & $T'_{peak}$ &  $S_G$ & $S_T$\\
 \midrule
 \multicolumn{2}{l}{Homogeneous distribution} & $ 1.30$ & $1.20$ & $2.5 \times 10^{-2}$ & $5 \times 10^{-2} $ \\
 $0.3$ & $0.15$ & $1,49$ & $1.29$ &  $1.7 \times 10^{-2}$ & $ 3.8 \times 10^{-2} $ \\
 $0.3$ & $0.10$ & $1.63$ & $1.35$ & $2.0 \times 10^{-2}$ & $ 4.3 \times 10^{-2} $ \\
 $0.6$ & $0.10$ & $3.50$ & $2.20$ & $4.2 \times 10^{-2}$ & $ 8.0 \times 10^{-2} $ \\
 $0.7$ & $0.15$ & $1.90$ & $1.65$ & $4.0\times 10^{-2}$ & $  7.5 \times 10^{-2} $ \\
 $0.7$ & $0.10$ & $2.35$ & $1.95$ & $4.9 \times 10^{-2}$ & $  9 \times 10^{-2} $ \\
 \bottomrule
\end{tabular}
\end{center}
\vspace{-0.5\baselineskip}
\captionof{table}{Values of interest as a function of varying particles distributions.}
\label{tab:inhomog}
%
% The position of the preferential radius of particles has a strong impact on both $G'$ and $T'$. 
% Figure \ref{grsd} shows the ratio of growth for several values of $\sigma$ with $r_d=0.3$, 
% with all other parameters being kept equal to those of Figure \ref{unird1000}.
% The effect of $\sigma$ is similar to the one observed for $r_d=0.3$. 
% On the other hand, the values of $G'_{peak}$, $T'_{peak}$ as well as $S_G$ and $S_T$ are quite 
% different. 
% For example, when $\sigma = 0.10$, 
% the peak of the ratio of the transient growth is $G'_{peak}(r_d=0.3) =1.65$ (compared to $G'_{peak}(r_d=0.7)=2.3$) 
% and the time ratio is $T'_{peak}(r_d=0.3) =1.35$ (compared to $T'_{peak}(r_d=0.7)=1.8$).
% The position of the particle distribution is therefore a critical parameter when considering the flow transient growth. 
% Although the effect is weaker, both $G'_{peak}$ and $T'_{peak}$ are significantly stronger for the 
% Gaussian particle distribution with $r_d=0.3$ than for a homogeneous particle distribution. 
% The Stokes number at which these peaks occur is smaller for Gaussian distribution when $r_d=0.30$ in 
% comparison to homogeneous distribution, this is the opposite effect to what occurred for $r_d=0.70$. 
% For $r_d=0.3$, $S_G = 2 \times 10^{-2}$ (while $S_G= 2.5 \times 10^{-2} $ with an homogeneous particle distribution) and
% $S_T \simeq 4 \times 10^{-2}$ (compared to $S_T=5 \times 10^{-2} $ for an homogeneous particle distribution).
%
\subsection{Influence of the radial distribution of particles}
Figure \ref{grsd} shows the ratios of gains as a function of $S$
for several average radii $r_d$ of the particle distribution, $Re=1000$ and $\sigma=0.1$. 
$G'_{peak}$ and $T'_{peak}$ show strong variations with $r_d$. 
Indeed, the $G'_{peak}$ ranges from $1.95$ to $3.50$ 
while $T'_{peak}$ ranges from $1.40$ to $2.20$.\\
The effect of particles on the flow is highest for $r_d$ in the range $0.5 - 0.6$ both for the ratio of 
maximum growths and the ratio of optimal times as shown in Figure \ref{grsd}. 
This value is relatively close to the Segr\'{e}-Silberberg radius where particles are known to naturally 
cluster \cite{segre1962behaviour}, albeit a little closer to the pipe centre.
Unlike the Reynolds number, $r_d$ has a very strong influence on the optimal Stokes number. 
Both $S_G$ and $S_T$ are larger than their counterpart in the case of a uniform particle distribution for all $S$, 
with the exception of $r_d=0.3$, which where this is only the case for $S\lesssim0.07$.
% However, the effect between $r_d$ and either $G'_{peak}$ or $T'_{peak}$ .
%
\section{Topology of the optimal perturbations}\label{sec:topology}

\begin{figure} 
\centering
\begin{subfigure}{0.50\textwidth}
\centering
\includegraphics[width=1\textwidth]{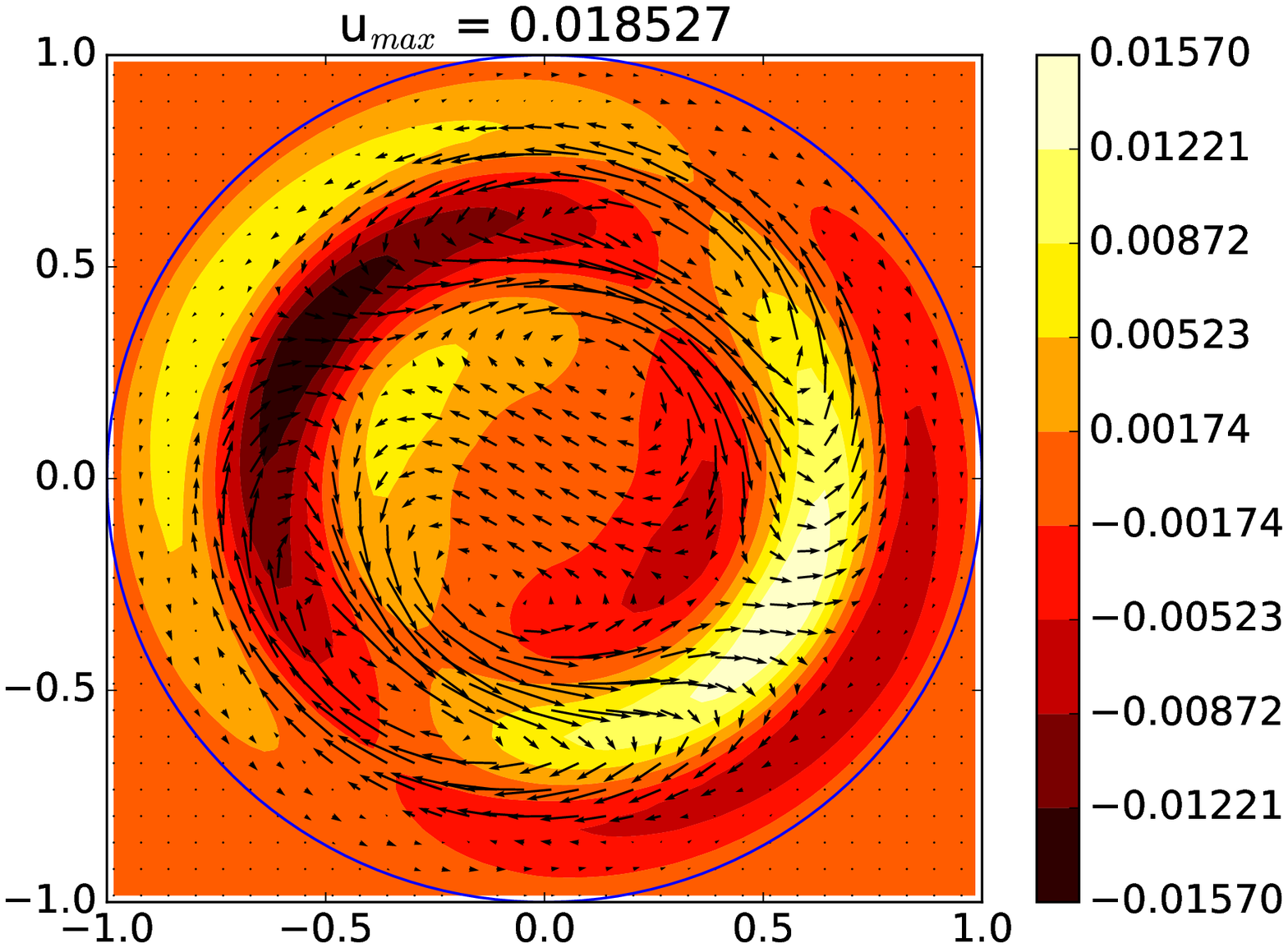}
\caption{Single phase flow, $T = 14$, $\mathbf{u_0}$ }
\label{fT14}
\end{subfigure}\hfill
\begin{subfigure}{0.50\textwidth} 
\centering
\includegraphics[width=1\textwidth]{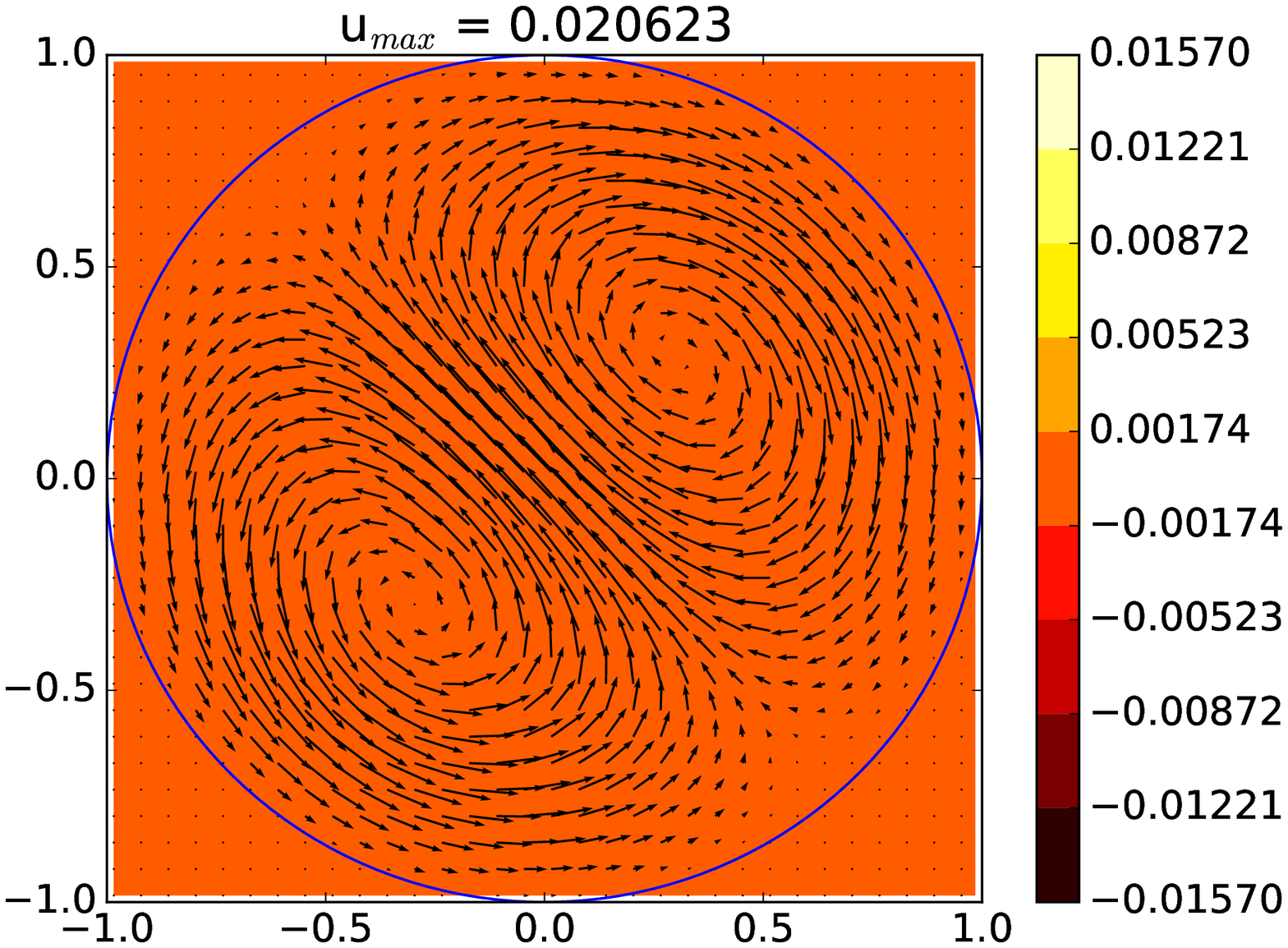}
\caption{Single phase flow, $T = 90$, $\mathbf{u_0}$}
\label{fT90}
\end{subfigure}
% \vskip\baselineskip
\begin{subfigure}{0.50\textwidth}
\centering
\includegraphics[width=1.0\textwidth]{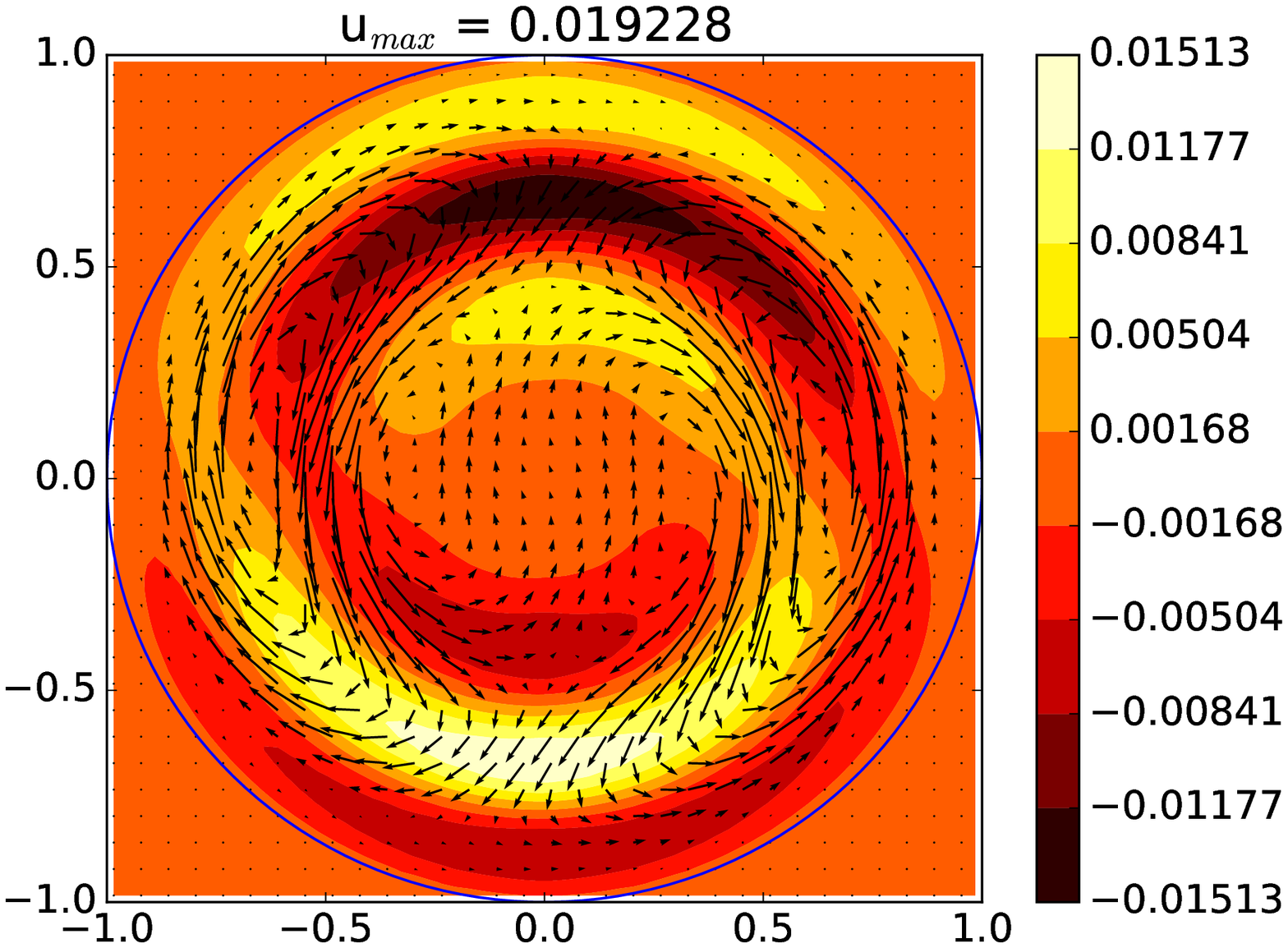}
\caption{Particulate flow, $T = 14$, $\mathbf{u_0}$ }
\label{gT14}
\end{subfigure}\hfill
\begin{subfigure}{0.50\textwidth}
\centering
\includegraphics[width=1.0\textwidth]{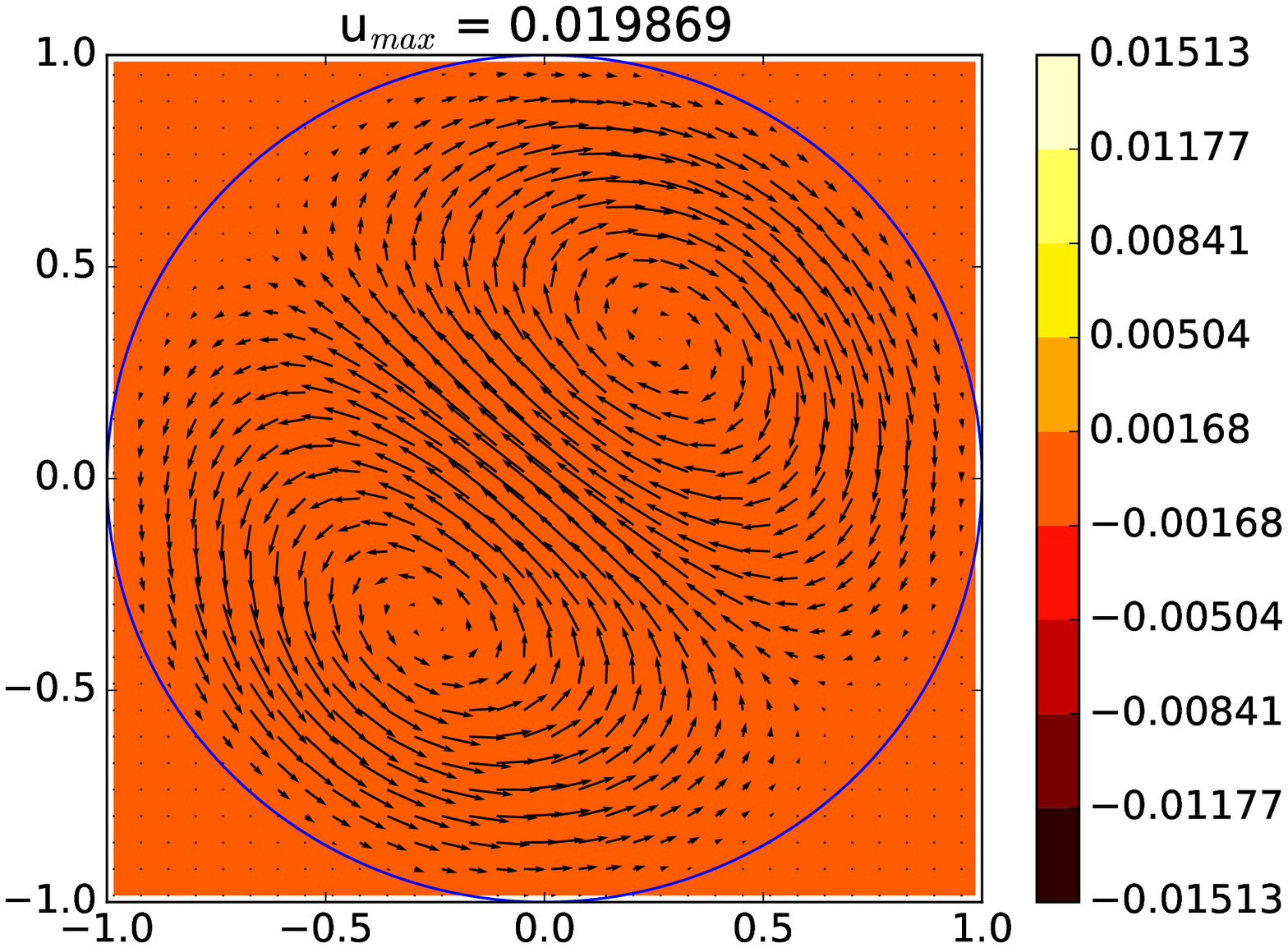}
\caption{Particulate flow, $T = 90$, $\mathbf{u_0}$ }
\label{gT90}
\end{subfigure}
% \vskip\baselineskip
\begin{subfigure}{0.50\textwidth} 
\centering
\includegraphics[width=1.0\textwidth]{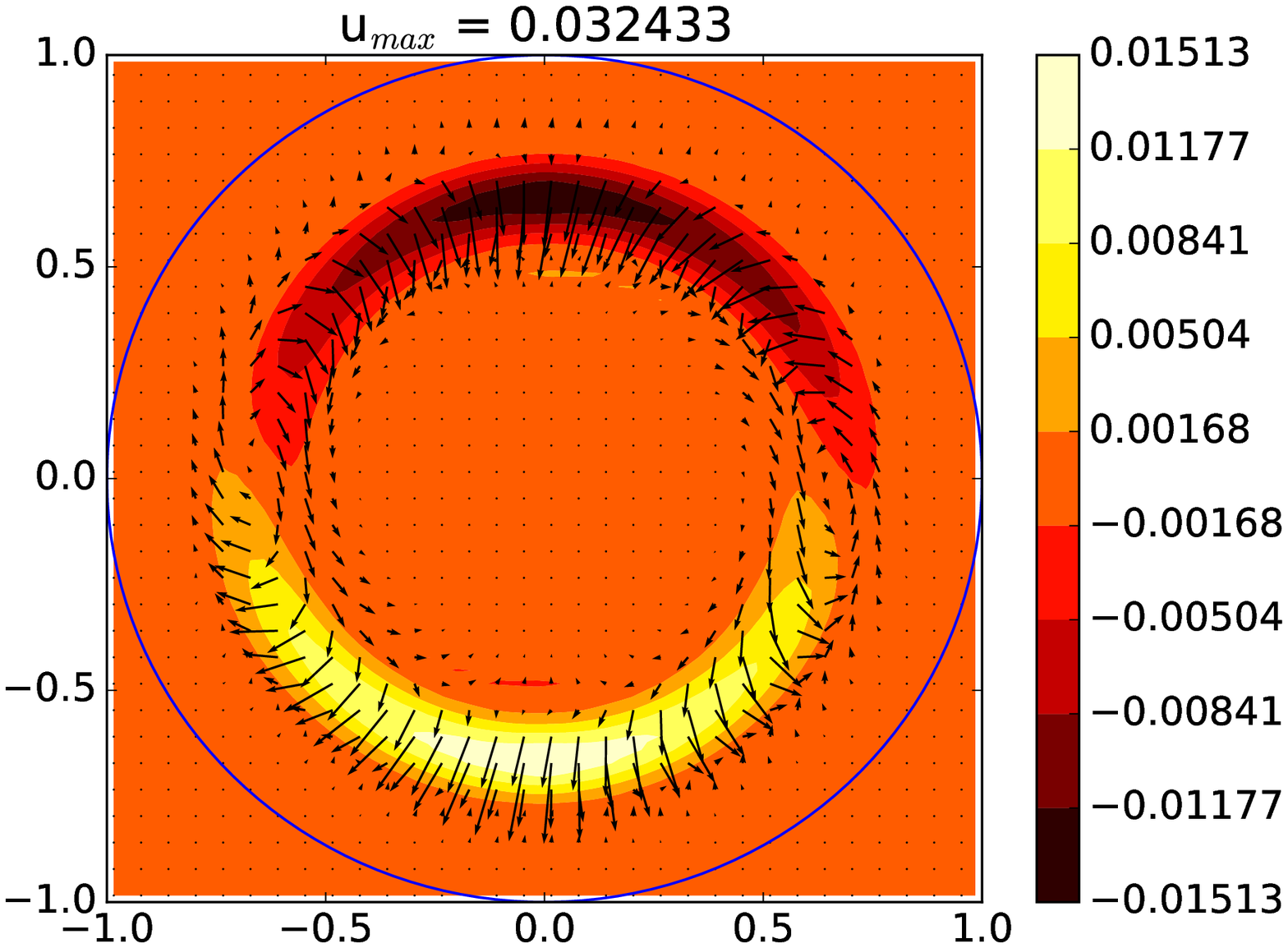}
\caption{Particulate flow, $T = 14$, $\mathbf{u_{p0}}$ }
\label{gpT14}
\end{subfigure}\hfill
\begin{subfigure}{0.50\textwidth} 
\centering
\includegraphics[width=1.0\textwidth]{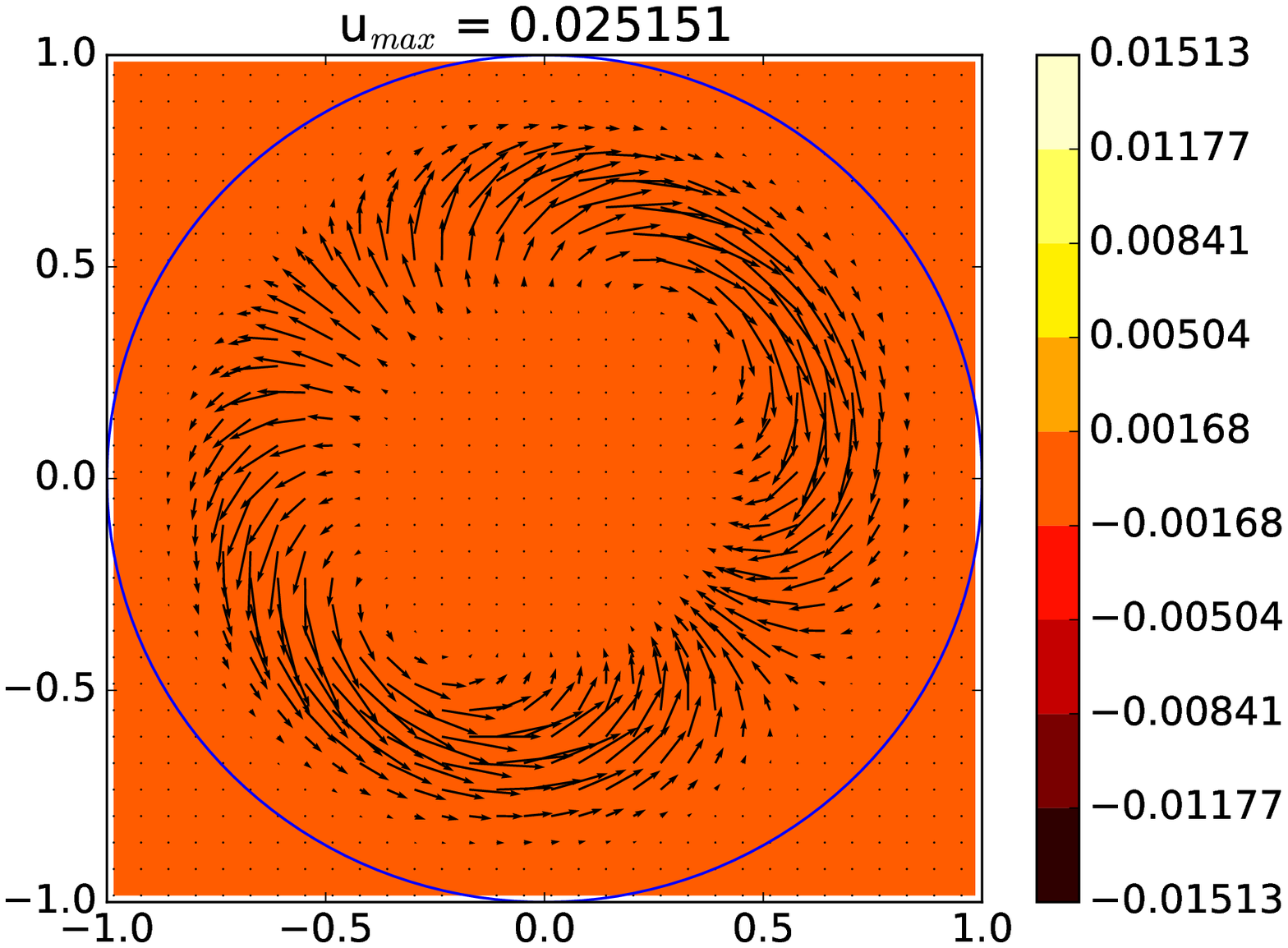}
\caption{Particulate flow, $T = 90$, $\mathbf{u_{p0}}$ }
\label{gpT90}
\end{subfigure}
\caption{Velocity contours of the optimal perturbation of a single phase flow and of a particulate flow with a Gaussian particle distribution, 
$Re=1500$, $f=0.1$, $S=10^{-3}$, $r_d=0.65$, $\sigma=0.104$.}
\end{figure}

In this section we study the topology of the optimal velocity fields. 
First, we consider the velocity fields at $t=0$, subsequently called optimal perturbation  
and denoted $\mathbf{u_{0}}$ for the fluid and $\mathbf{u_{p0}}$ for the particles' velocity. 
Second, the velocity fields at $t=T$, referred to as the velocity peak and denoted $\mathbf{u_{T}}$ for the fluid and $\mathbf{u_{pT}}$ for the particles' velocity.
Two target times are shown here, $T = 14$ for which the mode ($\alpha,m$) = ($1,1$) is dominant, 
and $T=90$ for which ($\alpha,m$) = ($0,1$) is dominant.
The radial sections of the optimal perturbations are very different depending 
on whether the dominant mode is ($\alpha,m$) = ($1,1$) or ($\alpha,m$) = ($0,1$).
Figures \ref{fT14}-\ref{gpT90} shows the contours of streamwise velocity and sectional fluid velocity vectors, 
for a single phase flow and a particulate flow with a nonhomogeneous distribution. 
 The optimal perturbations contours are weakly affected by the addition of particles in this case. 
%  The distinction is however more pronounced, for inhomogeneous particles distributions, for the particles optimal perturbation which 
%  velocities tend to be higher where particles are concentrated. 
For $T=14$ (figures \ref{fT14} and \ref{gT14}), the profile of the optimal perturbations shows two symmetric rolls in the spanwise direction.
The streamwise velocity has a peak in the shape of an antisymmetric annulus between $r=0.5$ and $r=0.7$.
Streamwise and spanwise velocities are of the same order of magnitude in both cases.
For a larger target time $T$, the streamwise independent mode is dominating 
as illustrated in figures \ref{fT90} and \ref{gT90}. 
In the spanwise direction we observe two rolls that are distinctive of the usual single phase transient growth.
Figures \ref{gpT14} and \ref{gpT90} shows the contours of streamwise velocity and sectional particles velocity vectors: both the streamwise and sectional particles velocities are strongest in the region where the 
particle concentration is the highest. 
\begin{figure}
\centering
\begin{subfigure}{0.50\textwidth}
\centering
\includegraphics[width=0.9\textwidth]{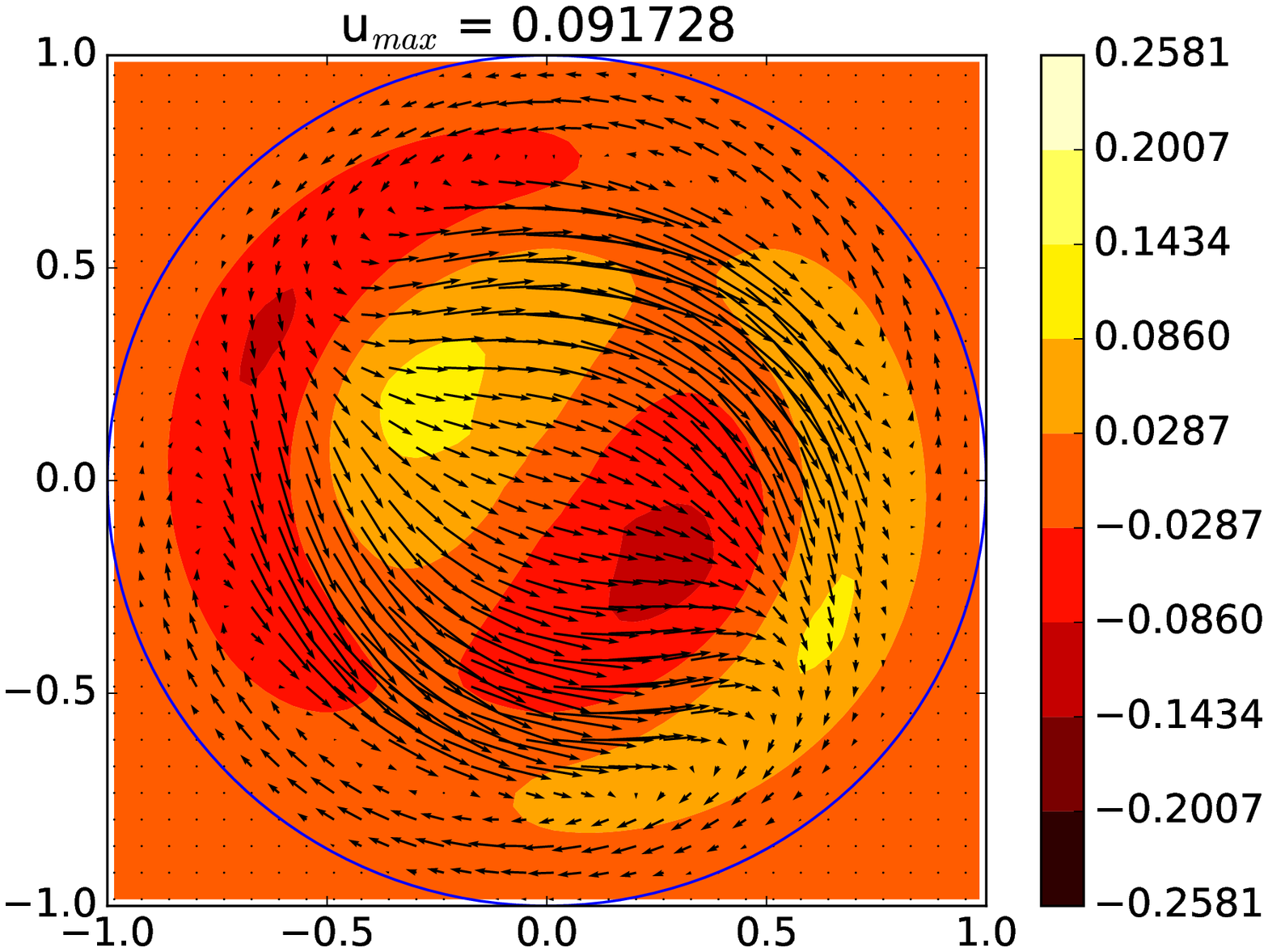}
\caption{Single phase flow, $T = 14$, $\mathbf{u_T}$}
\label{fT14f}
\end{subfigure}\hfill
\begin{subfigure}{0.50\textwidth} 
\centering
\includegraphics[width=0.9\textwidth]{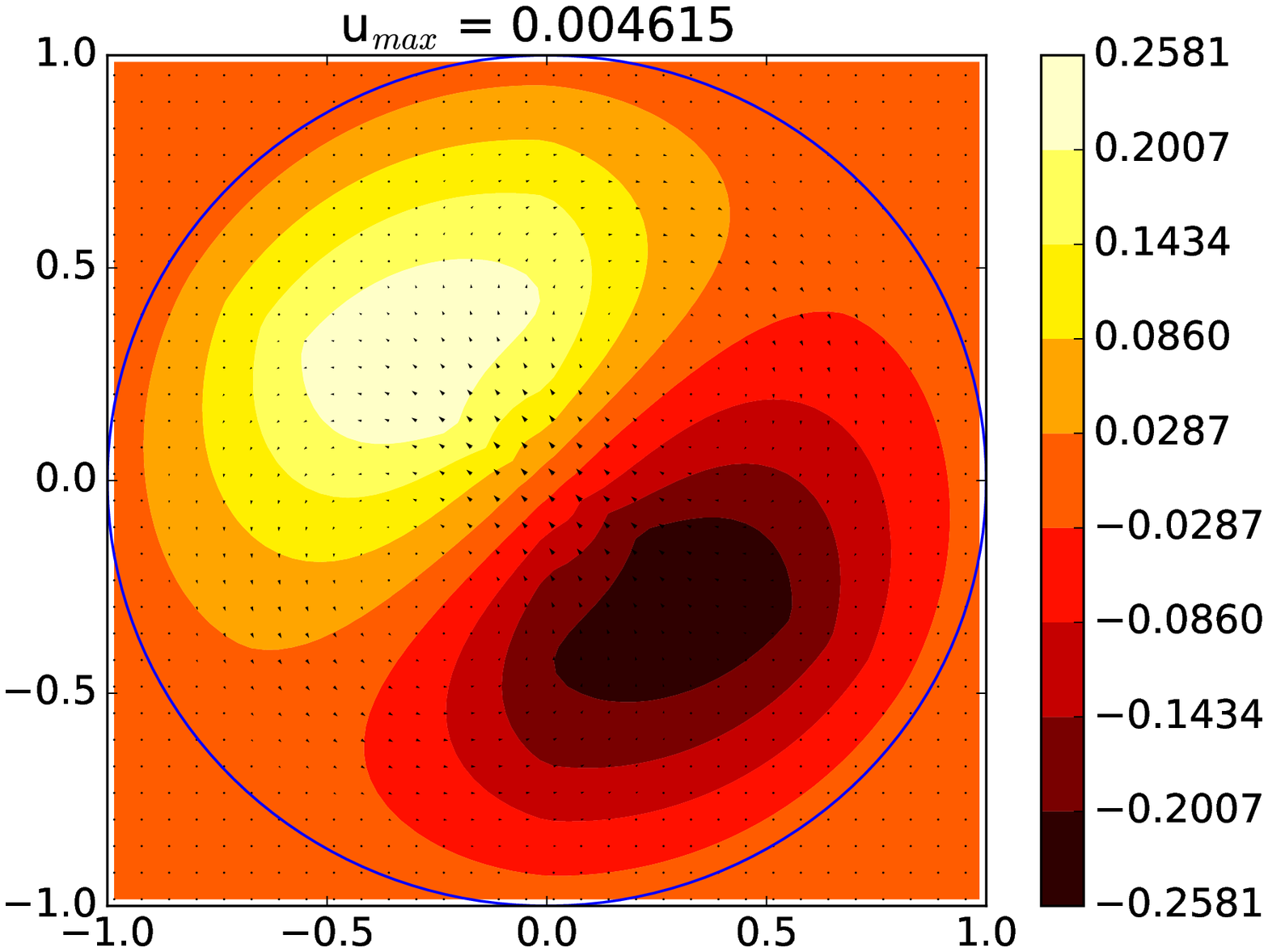}
\caption{Single phase flow, $T = 90$, $\mathbf{u_T}$}
\label{fT90f}
\end{subfigure}
% \caption{Peak velocity profiles for single phase flows.}
\begin{subfigure}{0.50\textwidth}
\centering
\includegraphics[width=0.9\textwidth]{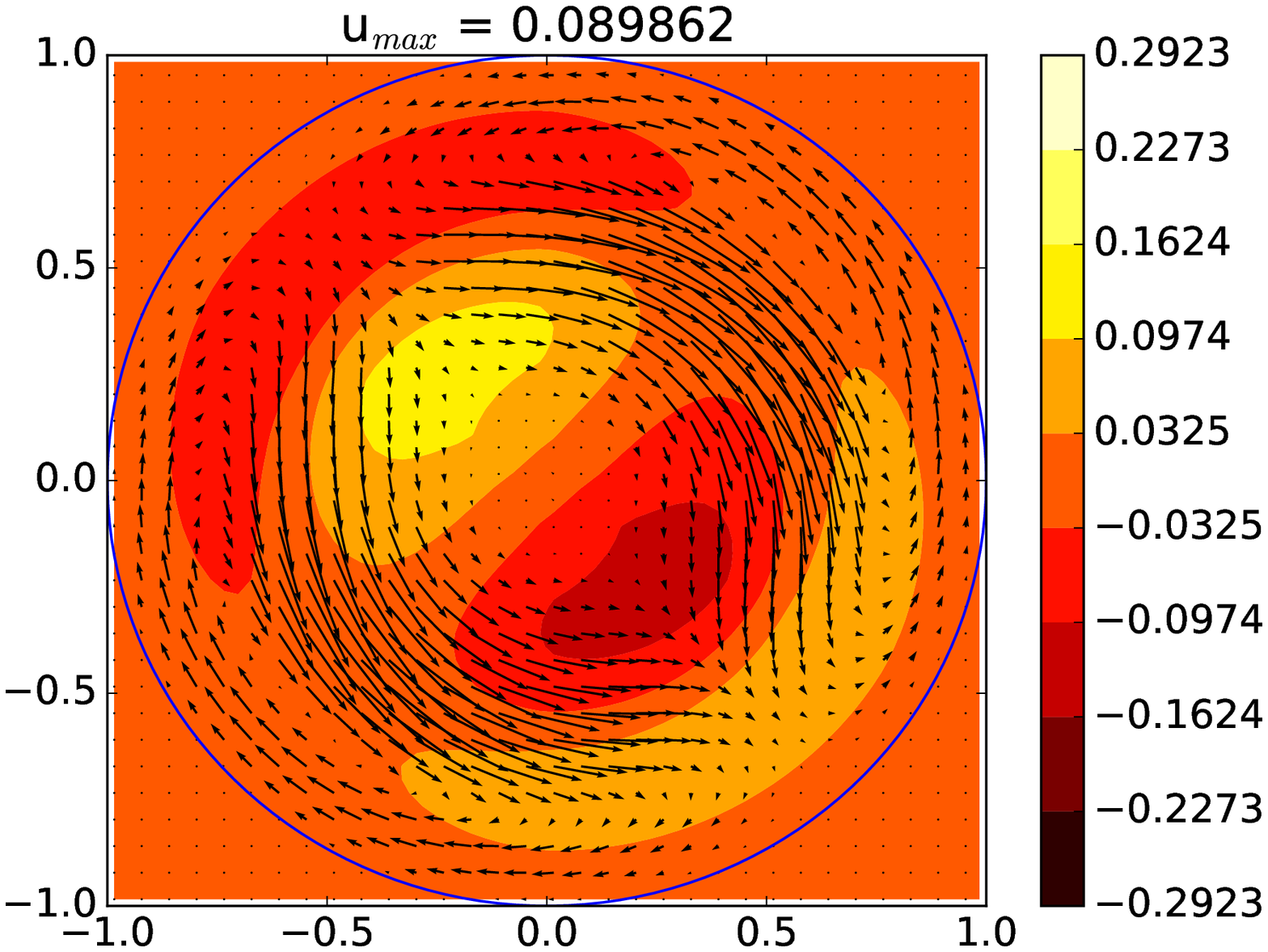}
\caption{Particulate flow, $T = 14$, $\mathbf{u_T}$ }
\label{gT14f}
\end{subfigure}\hfill
\begin{subfigure}{0.50\textwidth} 
\centering
\includegraphics[width=0.9\textwidth]{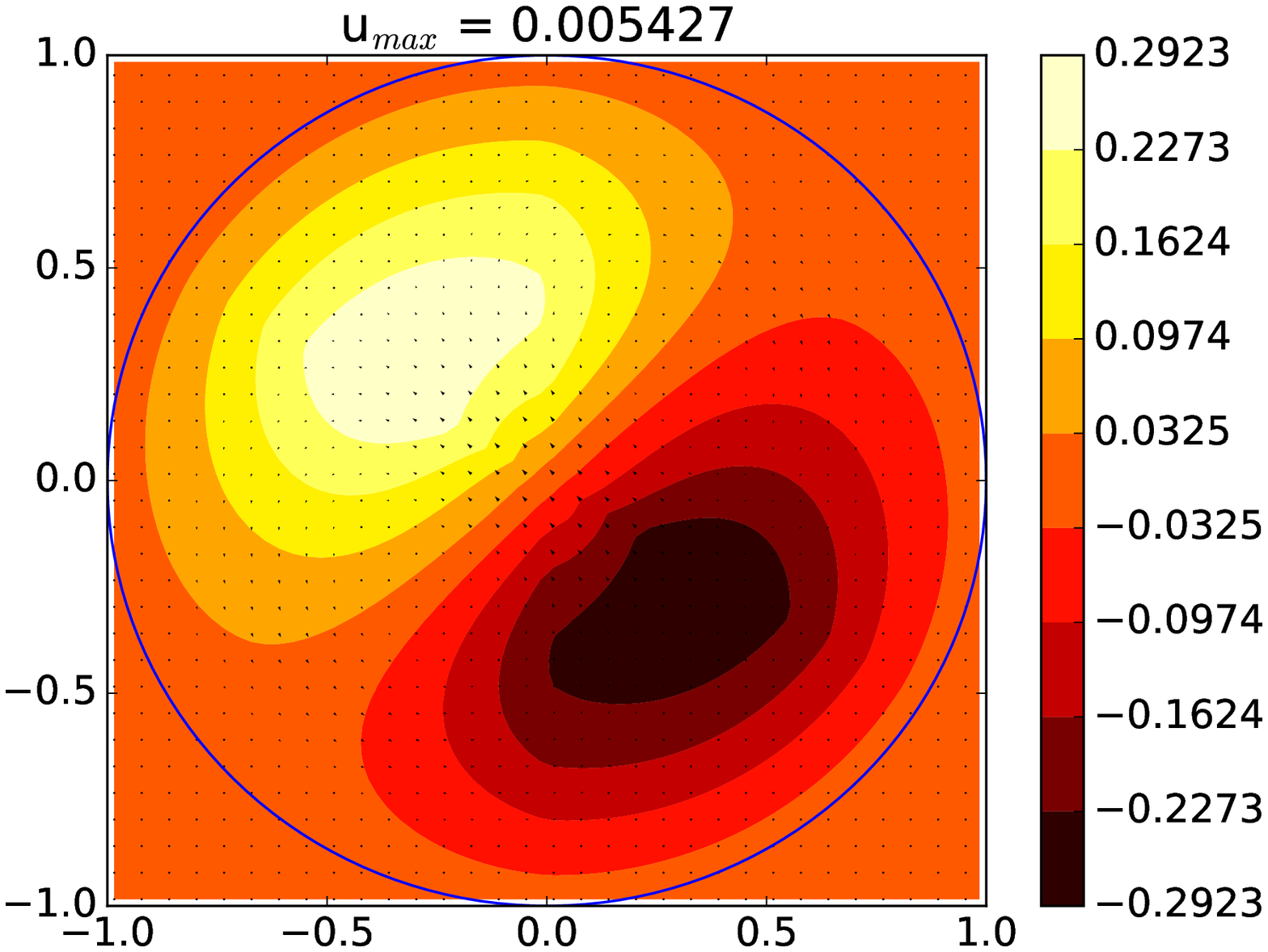}
\caption{Particulate flow, $T = 90$, $\mathbf{u_T}$ }
\label{gT90f}
\end{subfigure}
\centering
\begin{subfigure}{0.50\textwidth}
\centering
\includegraphics[width=0.9\textwidth]{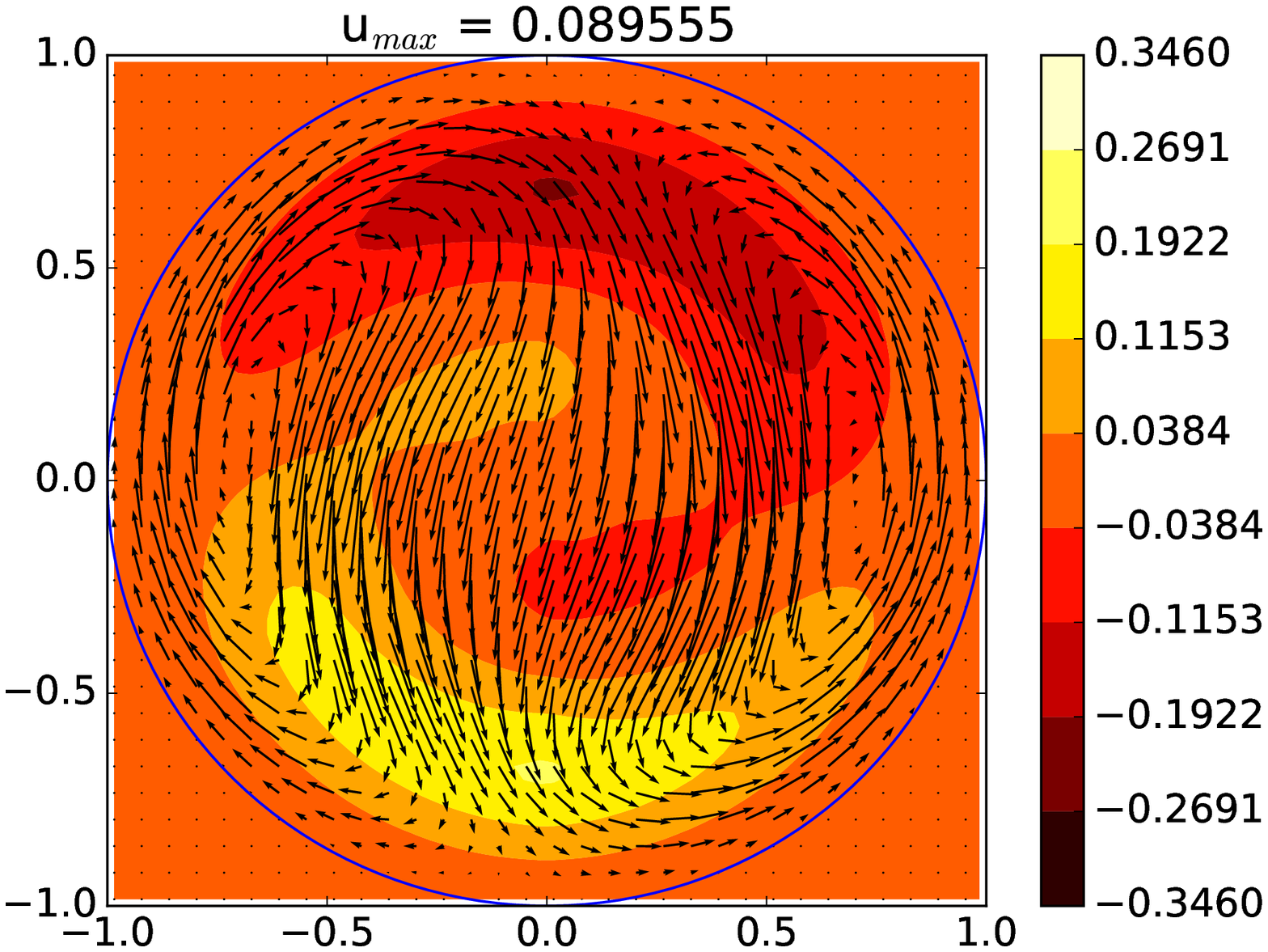}
\caption{Particulate flow, $T = 14$, $\mathbf{u_{pT}}$ }
\label{gTp14f}
\end{subfigure}\hfill
\begin{subfigure}{0.50\textwidth} 
\centering
\includegraphics[width=0.9\textwidth]{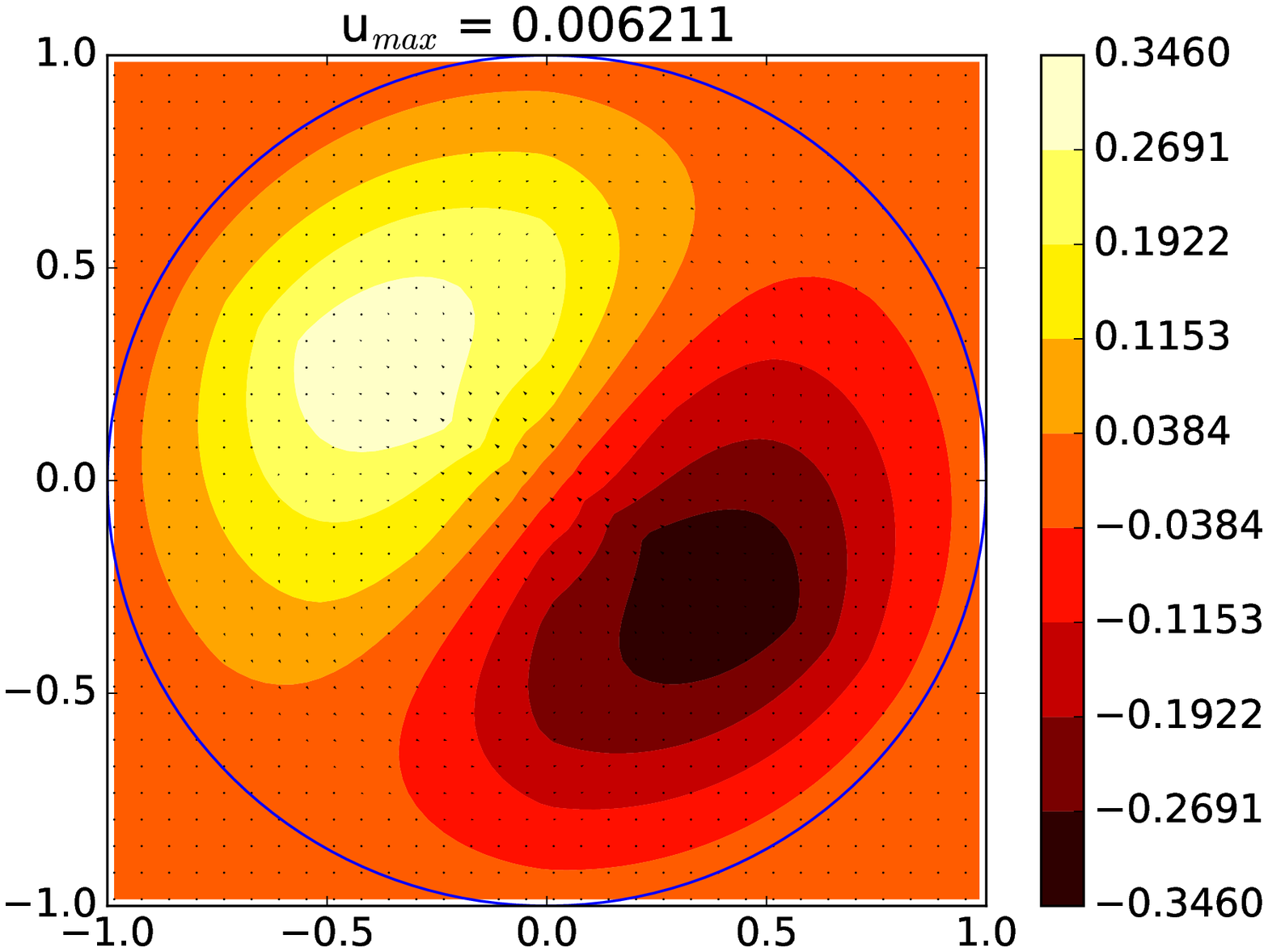}
\caption{Particulate flow, $T = 90$, $\mathbf{u_{pT}}$ }
\label{gpT90f}
\end{subfigure}
\caption{Peak velocity contours of single phase flow and a particulate flow with a Gaussian particle distribution at $t=0$, 
$Re=1500$, $f=0.1$, $S=10^{-3}$, $r_d=0.65$, $\sigma=0.104$.}
\label{fig:peak}
\end{figure}
Examples of peak velocity contours are given in figure \ref{fig:peak}. 
In all cases considered the streamwise velocity is larger than the spanwise velocity, for both $\mathbf{u_{T}}$ and $\mathbf{u_{pT}}$.
The effect is even more pronounced for $T=90$. 
$u_T(=14)$ spanwise component is similar to $u_0(T=14)$'s but the streamwise component is different 
with two opposed currents, one close to the centre and the other around it. 
In the case $T=90$ the streamwise velocity take the form of two antisymmetric rolls.
% The velocity contours are again not significantly affected by the addition of the particles, 
% and as opposed to what we have observed for the optimal perturbation, 
Figure \ref{fig:peak} shows that fluid and particles 
profile are, in the case of the peak velocity, almost identical, due to the 
strong coupling between the fluid and solid phases.\\
In summary, the addition of particles does not alter the transient growth mechanisms nor the topology 
of the optimal modes, even if the particles are distributed inhomohenously. Particles tend to be 
accelerated where the flow is, however their effect on the growth itself is significant when they 
are inhomogenously distributed.  

\newpage

\section{Conclusion and discussion}

This paper presented a study of the linear transient growth of particulate pipe flow through a simple 
two-fluid, two-way model for the solid and the liquid phases. 
The addition of particles has been found to increase the amount of transient growth regardless 
of Stokes number. However the modes that are responsible for the transient growth remain the same as 
those in flows without particles.  
The growth itself still varies with the Stokes number, with a sweet spot for which it is maximised. 
Interestingly, the corresponding Stokes number is independent of the Reynolds number. 
Moreover, the ratio of growths optimised over time for the particulate to non-particulate flow, 
$G'_{peak}$,  
is also independent of the Reynolds number, implying that the growth for the particulate 
flows scales as $Re^2$ as it does for the single phase flow \cite{bergstrom1993optimal}.

We also showed that the solid phase has a delaying effect on the transient growth, again 
regardless of the Stokes number considered. 
We observe that there is a value of $S$ for which the delay is maximised, 
this Stokes number is independent of the Reynolds number as well. 
Here too, the ratio of times of optimal growths for the particulate to the non-particulate flow 
$T'_{peak}$, is independent of the Reynolds number. This implies that 
the time for which growth is optimised scales as $Re$ as it does for the single phase flow.\\
The most important result is that allowing for particles to be inhomogeneously distributed 
can drastically increase their impact on the transient growth, which 
is increased by more than $200\%$ depending on their size and the shape of the spatial distribution. 
The way in which the particles are distributed is important too. We have considered particles in 
a Gaussian distribution of standard deviation $\sigma$ located in annulus located at radius $r_d$.
The transient growth increases monotonically as $\sigma$ decreases, \emph{i.e.} when the particles are 
more localised. The effect of the solid phase on the transient growth was also found to be weaker when 
the particles are localised close to the wall ($r_d$ close to 1) or at the pipe centre ($r_d$ close to 0), and strongest in the intermediate region ($r_d = 0.6-0.7$).  
This region seems to play a particularly important role both in the laminar state and in the 
process of transition to turbulence: not only do particle tend to naturally cluster there in the 
laminar state \citep{segre1962behaviour,matas2004inertial}, but particulate pipe flows have been 
found linearly unstable when particles of intermediate size are added in that region 
\cite{rouquier2018instability}. This raises the question of whether the actual pathway to turbulence 
is indeed sensitive to particles being present in that region, a question that could be answered in 
further analysis including fully nonlinear effects. Secondly, the question of the robustness of the 
model is also crucial: further studies with more physically accurate model for the solid phase and 
their interaction with the fluid phase are needed to confirm the nature of the role played by the 
mechanisms identified in this work.\\

%Although the addition of particles does increase the amount of transient growth of the flow, 
%Experimentally, particles have been found to, for certain sizes and volume concentrations,  
%lower the critical Reynolds number, although the mechanism 
%is exact mechanism is unclear \citep{matas2003transition}. 
%The fact than this was not observed in this study is likely due to either the linear transient not being able 
%to capture the dynamics of transition to turbulence or that the two-fluid model used for the description of the solid phase is 
%valid in a parameter range different from the one studied in \citep{matas2003transition} 
%(very small, heavy particles in a dilute against larger, neutrally buoyant particles with densely concentrated. 
%Follow-up research including nonlinear effect and a description of the solid phase 
%more suited to the problem would be necessary to capture the underlaying mechanisms 
%leading to transition to turbulence for particulate pipe flows.

\textit{AR is supported by TUV-NEL. CCTP is partially supported by EPSRC grant No. EP/P021352/1. AP acknowledges 
support from the Royal Society under the Wolfson Research Merit Award Scheme (Grant WM140032).}

\bibliographystyle{vancouver}
\bibliography{bibliography_thesis}%You need a file 'literature.bib' for this.

% \appendix

% \section{Derivation of the adjoint system of equations}

\end{document}